\documentclass{article}
\usepackage[utf8]{inputenc}
\usepackage[framed,numbered,autolinebreaks,useliterate]{mcode}
\usepackage{mathtools}
\usepackage{diffcoeff}
\usepackage[document]{ragged2e}
\usepackage{float}
\usepackage{physics}
\usepackage{amsmath}
\usepackage{amssymb}
\usepackage{tikz}
\usetikzlibrary{shapes,arrows}
\usepackage{geometry}
\usepackage{graphics}
\usepackage{caption}
\usepackage{subcaption}
\usepackage{placeins}
\usepackage{titlesec}
\newlength{\thickarrayrulewidth}
\titleformat{\section} 
	{\normalfont\Large\bfseries}{\makebox[30pt][l]{\thesection}}{0pt}{} 
\titleformat{\subsection} 
	{\normalfont\large\bfseries}{\makebox[30pt][l]{\thesubsection}}{0pt}{}
\titleformat{\subsubsection} 
	{\normalfont\normalsize\bfseries}{\makebox[40pt][l]{\thesubsubsection}}{0pt}{}

\usepackage{graphicx} 
\graphicspath{ {./Images/} }
\usepackage[rightcaption]{sidecap}
\usepackage{wrapfig}
\usepackage{listings}
\usepackage{xcolor}

\definecolor{codegreen}{rgb}{0,0.6,0}
\definecolor{codegray}{rgb}{0.5,0.5,0.5}
\definecolor{codepurple}{rgb}{0.58,0,0.82}
\definecolor{backcolour}{rgb}{0.95,0.95,0.92}

\begin{document}
\begin{center}
{\huge \bfseries Integral equation method for microseismic wavefield modelling in anisotropic elastic media}\\[0.3cm]\par

{\Large \bfseries Ujjwal Shekhar$^{1,a}$, Morten Jakobsen$^{1,a}$, Einar Iversen$^{1,a}$, Inga Berre$^{1,b}$, Florin A. Radu$^{1,b}$\\
(January 10, 2023)\\
  \large \bfseries$^1$ Center for Modeling of Coupled Subsurface Dynamics, University of Bergen, Bergen 5007, Norway\\
  $^a$ Department of Earth Science, University of Bergen\\
  $^b$ Department of Mathematics, University of Bergen\\
 Email: Ujjwal.Shekhar@uib.no}
\end{center}\par

\section*{ABSTRACT}
In this paper, we present a frequency-domain volume integral method to model the microseismic wavefield in heterogeneous anisotropic-elastic media. The elastic wave equation is written as an integral equation of the Lippmann-Schwinger type, and the seismic source is represented as a general moment tensor. The displacement field due to a moment tensor source can be computed using the spatial derivative of the elastodynamic Green's function. The existing matrix-based implementation of the integral equation is computationally inefficient to model the wavefield in a three-dimensional earth. An integral equation for the particle displacement is, hence, formulated in a matrix-free manner through the application of the Fourier transform. The biconjugate gradient stabilized method is used to iteratively obtain the solution of this equation. We apply the numerical scheme to three different models in order of increasing geological complexity and obtain the elastic displacement fields corresponding to the different types of moment tensor sources. The volume integral method has an advantage over the time domain methods in regard to adding multiple sources since it can work with discrete frequencies, one by one, and limit the computational cost. The generated synthetic data can be useful in inversion for the microseismic source and model parameters. \\[4pt]

\textbf{Keywords}: Integral equation, microseismic, moment tensor, seismic anisotropy.

\section{INTRODUCTION}

Activities such as fluid injection in the subsurface and hydraulic fracturing may lead to induced earthquakes and microseismicity. Microseismic wavefield modelling is one of the key aspects of a microseismic monitoring project. The source in microseismic events cannot be represented by a vectorial point source. Instead, moment tensors have been used to describe the source mechanisms of earthquakes (Aki and Richards, 2002; Jost and Herrmann, 1989). Several authors (Graves, 1996; Shi et al., 2018; Lei et al., 2021) have performed seismic full-waveform modelling in the time domain with moment tensor source representation. They employed the staggered-grid finite difference method (Virieux, 1986) to solve the elastodynamic equation in the velocity-stress formulation. To obtain a symmetric moment tensor source, they evenly distributed the shear stress increments on the adjacent shear-stress grid points around the true moment tensor source location. In this research, we apply a volume integral method to model the microseismic wavefield in the frequency domain. The moment tensor source can be introduced at any grid point by using an equivalent body force term. \\[4pt]

The foundations of integral equations lie in the general principles of scattering theory. In the integral equation method, the actual medium is decomposed into a homogeneous reference or background medium and a contrast medium. It is only required to discretize the domain where the scattering potential is nonzero when using the volume integral method, whereas the finite difference method requires a full discretization of the model. Using Green's function for the background medium and numerical values of the physical properties in the contrast medium, we obtain the wavefield in the actual medium. The integral equation approach has been widely applied to solve the nonlinear scattering problem in electromagnetics (Jakobsen and Tveit, 2018). The same methodology can easily be extended to elastic wave scattering problems. Gibson and Ben-Menahem (1991) clearly demonstrated the application of elastic wave scattering theory to study wave propagations in fractured media. The Born approximation (Miles, 1960; Hudson and Heritage, 1982) to a fully elastic scattering obstacle is one of the frequently applied methods to compute the scattered field in elastodynamic wave propagation. It is used to calculate the wavefields of Rayleigh scatterings due to the perturbations to elastic constants considering only a single scattering of elastic waves. However, the iterative solver-based volume integral method accounts for multiple scatterings of elastic waves. The Born approximation also predicts incorrect travel times, whereas the full integral equation method accounts for the travel time difference between the actual medium and the background medium. \\[4pt]

Within the framework of seismic wavefield modelling, a limited number of works have been concluded using the volume integral method, and most of those focused on acoustic waves only (Alles and van Dongen, 2011; Malovichko et al., 2018; Jakobsen and Ursin, 2015; Xiang et al., 2021). An acoustic or elastic wave equation can be written as a corresponding integral equation of the Lippmann-Schwinger type, and different numerical techniques can be employed to solve this equation. Jakobsen and Ursin (2015) used the transition operator approach for seismic forward modelling in the acoustic approximation. Malovichko et al. (2018) presented frequency-domain acoustic 3D modelling using the integral equations method and proved that the method can accurately calculate the pressure field for models with sharp material boundaries. Tong and Chew (2009) developed a multilevel fast multipole algorithm as an accelerator for integral equation solvers for elastic wave scattering problems. Touhei (2011) proposed a fast method for the volume integral equation for elastic wave propagation in a half space. The fast transform was implemented by decomposing the kernel of the transform into the ordinary Fourier and Laplace transforms. \\[4pt]

Previous applications of the volume integral equation method (Touhei, 2011) to elastic wavefield modelling have ignored the effects of seismic anisotropy and have not presented synthetic seismograms for realistic seismic models. In the volume integral method, the heterogeneity and anisotropy parameters can be assigned directly, however attaining the solution of the resulting integral equation is a challenging task. An effort to incorporate the parameters for the arbitrary anisotropic media and then solving the system of equations was made by Jakobsen et al. (2020). They used a matrix-based solver for the two coupled integral equations to obtain the elastic wavefield. Although they generalized their method to heterogeneous-anisotropic media, the numerical examples presented did not consider a variable density medium. Moreover, the matrix-based method used to solve the integral equation was inefficient in terms of computational cost and memory. \\[4pt]

Microseismic events were not taken into account in previous works on the integral equation method for elastic wave scattering problems. Our goal is to carry out microseismic wavefield modelling using the volume integral method. To simulate the wavefield in reservoir-scale 3D models, we focus on the development of a computationally efficient numerical solver for the integral equation. In Section 2, we show how the integral equation can be formulated to incorporate the moment tensor source into the heterogeneous anisotropic-elastic media. We utilize the Lippmann--Schwinger equation for elastic waves (Jakobsen et al., 2020) and implement the moment tensor source (Stein and Wysession, 2003; Madariaga, 2015) in the equation using a body force term. In Section 3, we demonstrate how to efficiently solve the integral equation in the particle displacement. We emphasize the Fourier transform-based iterative solver to perform the modelling work in a matrix-free manner. Symmetries in the off-diagonal components of the moment tensor, the stiffness tensor and the strain tensor allow for the use of the Voigt notation during the implementation. In the integral equation method, components of the strain field are also needed to obtain the particle displacement components. The finite difference method is used to numerically compute the strain. In Section 4, we apply the method to a homogeneous three-dimensional earth model. We observe the response of the wavefields to different types of moment tensor sources. We next test the accuracy of the integral equation method in computing the microseismic wavefield in the anisotropic-elastic media. This is accomplished by comparing the integral equation result with the finite difference time domain result for the same model parameters. Finally, we examine the microseismic wavefield in laterally inhomogeneous media to explore the applicability of the integral equation method for heterogeneous models.

\section{INTEGRAL EQUATION FORMULATION}
Waves generated in heterogeneous media can be reflected, refracted, transmitted and/or diffracted. The elastodynamic wave equation takes all these phenomena into account. The components of the particle displacement vector $u_{i}({\bf x})$
at point ${\bf x}$ due to a force source density with components $f_{i}({\bf x})$ in an anisotropic elastic medium with
stiffness tensor component $c_{ijkl}({\bf x} )$ and mass density $ \rho ({\bf x})$ satisfies the elastodynamic wave equation
\begin{equation}
\left[c_{ijkl}({\bf x})u_{k,l}({\bf x, \omega} )\right]_{,j} + \rho ({\bf x})\omega ^{2}u_{i}({\bf x, \omega})= -f_{i}({\bf x, \omega} ),
\label{Eq1}
\end{equation} 
where $\omega $ is the angular frequency. The value of each of the indices can be 1, 2 and 3, indicating the X, Y and Z components, respectively. We assume that $u_{i}({\bf x})$ is a plane-wave solution to Equation (1). In the integral equation approach, the complex interactions of the wave phenomena are termed scatterings. In this method, we decompose the stiffness tensor components $c_{ijkl}({\bf x})$ and the mass density $\rho ({\bf x})$ as
\begin{equation}
 c_{ijkl}({\bf x} ) =  c^{(0)}_{ijkl}  + \Delta c_{ijkl}({\bf x} ) , 
 \label{Eq2}
\end{equation}
\begin{equation}
\rho ({\bf x}) = \rho ^{(0)} + \Delta \rho ({\bf x }), 
\label{Eq3}
\end{equation}
where $c^{(0)}_{ijkl}$ and $\rho ^{(0)}$ are the stiffness coefficients and mass density, respectively, of an arbitrary homogeneous reference/background medium and $\Delta c_{ijkl}({\bf x} ) $ and $\Delta \rho ({\bf x})$ are the corresponding contrast terms.
\\[4pt]
From Equations (1), (2) and (3), we obtain
\begin{equation}
\begin{split}
 \left[c_{ijkl}^{(0)}({\bf x})u_{k,l}({\bf x, \omega} )\right]_{,j} + \rho ({\bf x})\omega ^{2}u_{i}({\bf x, \omega}) & = -f_{i}({\bf x} ) -\left[ \Delta c_{ijkl}({\bf x})u_{k,l}({\bf x, \omega})\right] _{,j} \\ &  -  \omega ^{2}\Delta \rho ({\bf x})u_{i}({\bf x, \omega}). 
\label{Eq4}   
\end{split}
\end{equation}
\\
The second and third terms on the right-hand side of Equation (4) represent the so-called contrast sources of the moment tensor and force vector type, respectively. By treating the contrast sources similar to ordinary sources and using Green's function representation (Jakobsen et al., 2020), we obtain the following integral equation:
\begin{equation}
 u_{i}({\bf x} ) =  u^{(0)}_{i}({\bf x} ) + 
 \int_\Omega d{\bf x}'G_{ij}^{(0)}({\bf x}, {\bf x}' )\left \{ \left[\Delta c_{jklm}({\bf x}' ) u_{l,m}({\bf x}' )\right] _{,k}  
 +  \omega ^{2} \Delta \rho ({\bf x}')u_{i}({\bf x}') \right\}.
 \label{Eq5}
\end{equation}
\\
The integral equation in the particle displacement (Eqn. 5) demonstrates the linear superposition of the background and scattered fields. If the elastodynamic Green's functions $G_{ij}^{(0)}$ for the background model are known, it is only required to discretize the contrast model. The integral part has been computed over a scattered domain $\Omega$ and ${\bf x}'$ is a position vector within this region. The superscript (0) denotes a physical quantity in the homogeneous background medium. The wavefield in the homogeneous background model due to the vectorial source with components $f_{j}({\bf s} )$ is obtained by the following convolution integral:
\begin{equation}
 u_{i}^{(0)}({\bf x} ) = \int d{\bf x}G^{(0)}_{ij}({\bf x}, {\bf s} )f_{j}({\bf s} ),
 \label{Eq6}
\end{equation}
where $G_{ij}^{(0)}({\bf x}, {\bf s} )$ is the solution of
\begin{equation}
\left[c_{ijkl}^{(0)}G^{(0)}_{kn,l} ({\bf x}, {\bf s})\right]_{,j} + \rho \omega ^{2}G^{(0)}_{in}({\bf x},{\bf s})= -\delta _{in}\delta ({\bf x} -{\bf s}).
\label{Eq7}
\end{equation}
\\
Here, $\delta ({\bf x} -{\bf s})$ is the 3-D Dirac-delta function and $\delta_{in}$ is the Kronecker delta defined by $\delta_{in} = 1$ when $i = n$, and $\delta_{in} = 0$ when $i \neq n$. In elasticity theory, Green's function gives the displacement generated by a point force in a certain direction (Snieder, 2002). Here, Green's tensor $G^{(0)}_{ij}({\bf x}, {\bf s})$ is the displacement at location $\bf x$ in the $i$ direction due to the $i$th component of a unit force in the $j$ direction at location ${\bf s}$. The details on Green's function and its spatial derivative can be found in Appendix A.\\[4pt]

By performing a partial integration, and using the well-known symmetries of the elastic stiffness tensor, Equation (5) can be rewritten as (Červený, 2001)
\begin{equation}
 \begin{split}
u_{i}({\bf x}) & = u_{i}^{(0)}({\bf x}) + \int_\Omega d{\bf x}'H_{ijk}^{(0)}({\bf x},{\bf x}')\Delta c_{jklm}({\bf x}')\epsilon _{lm}({\bf x}') \\ & +  \omega ^{2}\int_\Omega d{\bf x}'G_{ij}^{(0)}({\bf x},{\bf x}' ) \Delta \rho ({\bf x}')u_{j}({\bf x}'),
\label{Eq8}     
 \end{split}
\end{equation}
where the strain tensor is the symmetric gradient of the displacement vector and is given by
\begin{equation}
 \epsilon _{kl}({\bf x} ) = \frac{1}{2}\left[u_{k,l} ({\bf x} ) + u_{l,k}({\bf x} ) \right]   
\end{equation}
and $H_{ijk}^{(0)}({\bf x}, {\bf x}')$ is the first-order spatial derivative of Green's tensor for the background medium. It represents the $i$th component of the elastic displacement due to the $jk$ component of the contrast stress tensor and is expressed as
\begin{equation}
H^{(0)}_{ijk}({\bf x}) = \frac{1}{2}\left[G_{ij,k}^{(0)}({\bf x}) + G_{ik,j}^{(0)}({\bf x})  \right] .
\label{Eq9}
\end{equation}
\\
Equation (8) is known as the Lippmann--Schwinger equation for elastic wave scattering. In the derivation of this integral equation, we have made use of the fact that Green's function for the background medium only depends on the difference between the source and receiver coordinates since a homogeneous background medium is translation invariant. We also used the fact that any tensor can be decomposed into symmetric and anti-symmetric parts, and that the contraction of a symmetric tensor with an anti-symmetric tensor is identical to zero (Pujol, 2003).\\[4pt]

Since the third-rank tensor field $H^{(0)}_{ijk}({\bf x})$ by construction is symmetric with respect to an interchange of the last two indices, we can express Equation (8) in an abbreviated notation as
\begin{equation}
\begin{split}
u_{i}({\bf x}) & = u_{i}^{(0)}({\bf x}) + \int_\Omega d{\bf x}'H_{iJ}^{(0)}({\bf x},{\bf x}')\Delta c_{JK}({\bf x}')\epsilon _{K}({\bf x}') \\
& +  \omega ^{2}\int_\Omega d{\bf x}'G_{ij}^{(0)}({\bf x},{\bf x}' ) \Delta \rho ({\bf x}')u_{j}({\bf x}').    
\end{split}
\label{Eq10}
\end{equation}
\\
Here, the small and capital indices take values from 1 to 3 and from 1 to 6, respectively, and Einstein's summation convention applies (see Auld, 1990). Finally, we note that the above equation can be expressed more compactly in the hybrid tensor-operator notation as
\begin{equation}
u_{i} = u_{i}^{(0)} + H_{iJ}^{(0)}\Delta c_{JK}\epsilon _{K} + \omega ^{2}G_{ij}^{(0)}\Delta \rho u_{j}.
\end{equation}
\\
In the above expression, we have kept the tensor indices and the abbreviated indices when expressing the integral equation in an  operator form. The use of hybrid tensor-operator notation is convenient when discussing the algorithms for solving the integral equation iteratively. To simplify the presentation, we use the same notation for the integral operator $H_{iJ}$ and its kernel function $H_{iJ}({\bf x},{\bf x}')$ defined in Equation (11).

\subsection{Moment tensor source representation}
A moment tensor can be used to represent the seismic source in the microseismic events related to the slip on a fault plane. The mathematical models of this type of source can be given by a 3 x 3 tensor $M_{ij}$. Due to the symmetry of the off-diagonal elements, it can be written in the Voigt notation as a 6 x 1 vector $M_{J}$. The expression of the moment tensor source can be understood by analysing the kinematics behind the distributions of the transformation strain. The transformation strain describes an alteration of the stress-free configuration of a solid (Rice, 1998). Depending on the focal mechanism for microseismic events, the moment tensor source can be categorized as a single dipole, double-couple, explosion-type source, etc. (Stein and Wysession, 2003). The moment tensor components of double-couple microseismic sources in anisotropic formations have been eloquently derived by Grechka (2020). \\[4pt]

The equivalent body force density (Aki and Richards, 2002) for the moment tensor source is
\begin{equation}
 f_{i}({\bf x}) = -M_{ij}\frac{\partial }{\partial x_{j}}\delta ({\bf x} - {\bf s})e^{i\omega t_{0}},   
\end{equation}
where $M_{ij}$ is a moment tensor component, ${\bf s}$ represents the positioning of the microseismic event and $t_{0}$ may be referred to as the rupture time or the timing of the microseismic event. Using Equation (13) in Equation (6), we obtain the background displacement field,
\begin{equation}
 u_{i}^{(0)}({\bf x}) = \int d{\bf x} G^{(0)}_{ij}({\bf x},{\bf s})M_{jk}\frac{\partial }{\partial x_{k}}\delta ({\bf x} - {\bf s})e^{i\omega t_{0}}.
\end{equation}
\\
It is a common practice to assume that the moment tensor component $M_{ij}$ is constant during the entire period of seismic activity. We can simulate the seismic radiation from a moment tensor source by using the spatial derivative of Green's function. We exploit the property of the Dirac-delta function; that is, if we integrate the product of a function and the derivative of the Dirac-delta function over the spatial domain, the result is the spatial derivative of that function computed at the source location. The particle displacement field at position ${\bf x}$ due to a moment tensor source at position ${\bf s}$ is now given by
\begin{equation}
 u_{i}^{(0)}({\bf x}) = \int d{\bf x} H_{ijk}^{(0)}({\bf x}, {\bf s})M_{jk}e^{i\omega t_{0}}.   
\end{equation}
\\
In an abbreviated notation by 
\begin{equation}
 u_{i}^{(0)}({\bf x}) = \int d{\bf x} H_{iJ}^{(0)}({\bf x}, {\bf s})M_{J}e^{i\omega t_{0}}.   
\end{equation}

\section{EFFICIENT IMPLEMENTATION OF THE INTEGRAL EQUATION METHOD}

We work in the frequency domain and want to demonstrate wave propagation in heterogeneous anisotropic media using the full integral equation solution. To develop a computationally efficient numerical scheme, we employ the following methodology:

\subsection{Finite difference for the strain field}

The strain tensor in Voigt notation can be written as
\begin{equation}
  \epsilon =  \begin{bmatrix}
\epsilon_{1} & \epsilon_{2} & \epsilon_{3} & \epsilon_{4} & \epsilon_{5} & \epsilon_{6} 
\end{bmatrix}^T , 
\end{equation} 
where $\epsilon_{1} \equiv \epsilon_{11}$, $\epsilon_{2} \equiv \epsilon_{22}$, $\epsilon_{3} \equiv \epsilon_{33}$, $\epsilon_{4} \equiv 2\epsilon_{23}$, $\epsilon_{5} \equiv 2\epsilon_{13}$ and $\epsilon_{6} \equiv 2\epsilon_{12}$.
\\[4pt]
In the integral equation formulation, there is a simultaneous computation of both the strain field and the displacement field. If we solve the two coupled integral equations for the particle displacement and strain (Jakobsen et al., 2020), the numerical computation is slow. This is because it requires the calculation of the second-order spatial derivative of the background Green's function. We compute the strain field using the finite difference operator on the elastic displacement vectors to calculate its derivative with respect to the position vector. Let the grid spacing in each of the directions be $h$ and the displacement components in the X, Y and Z directions be $u_1$, $u_2$ and $u_3$, respectively. The strain field components at a grid node $i$,$j$,$k$ are given as:
\begin{equation}
\begin{aligned}
\epsilon_{1} (i, j, k) &=  \frac{ u_1 (i + 1, j, k) - u_1 (i - 1, j, k)}{2h}\\
\epsilon_{2} (i, j, k) &=  \frac{ u_2 (i, j + 1, k) - u_2 (i, j - 1, k)}{2h}\\
\epsilon_{3} (i, j, k) &=  \frac{ u_3 (i, j, k + 1) - u_3 (i, j, k - 1)}{2h}\\
 \epsilon_{4} (i, j, k) &= \frac{1}{2h} \left[  u_2 (i, j, k + 1) - u_2 (i, j, k - 1) +  u_3 (i, j + 1, k) - u_3 (i, j - 1, k)\right]\\
 \epsilon_{5} (i, j, k) &= \frac{1}{2h} \left[ u_1 (i, j, k + 1) - u_1 (i, j, k - 1) +   u_3 (i + 1, j, k) - u_3 (i - 1, j, k)\right]\\   
 \epsilon_{6} (i, j, k) &= \frac{1}{2h} \left[  u_1 (i, j + 1, k) - u_1 (i, j - 1, k) +   u_2 (i + 1, j, k) - u_2 (i - 1, j, k)\right]\\    
\end{aligned}.
\end{equation}
\\
To compute the strain field near the model boundaries, depending on the position where we are computing the derivative, we use either the forward or the backward difference.

\subsection{FFT accelerated Krylov subspace method}

Modelling seismic data through the implementation of a matrix-based method involves high computational costs and memory consumption. We utilize an iterative solver based on the fast Fourier transform (Stein and Wysession, 2003) to perform efficient seismic modelling, i.e., in a matrix-free manner. The fast Fourier transform, or FFT iterative scheme, is based on a recurrence relation and does not lead to the resolution of a linear system (Monchiet and Bonnet, 2012). The computational cost decreases from the order of $N^3$ for the matrix-based direct solver method to the order of $NlogN$ for the matrix-free iterative method, where $N$ is the number of grid points. The computational memory scales down from the order of $N^2$ to the order of $N$.\\[4pt]

Equation (12) can be rearranged as
\begin{equation}
u_i - \left[ H^{(0)}_{iJ}\Delta c_{JK}\epsilon_K + \omega^2 G^{(0)}_{ij} \Delta \rho u_j \right] = u^{(0)}_i    
\end{equation}
or
\begin{equation}
\left[\delta _{il} - \frac{1}{2}H^{(0)}_{iJ}\Delta c_{JK}\nabla_{Kl} - \omega^2 G^{(0)}_{ij} \Delta \rho\delta _{jl}\right]u_l = u^{(0)}_i.    
\end{equation}
\\
A fast formulation of the displacement field can be obtained using the background Green's function and the spatial derivative of the background Green's function in the Fourier transformed coordinates (k-space):
\begin{equation} 
 \begin{split}
u_i({\bf x}) - \mathcal{F}^{-1}\left[H^{(0)}_{iJ}({\bf k})\circ\mathcal{F}[\Delta c_{JK}({\bf x})\epsilon_K({\bf x})]\right] & - \\ \omega^2 \mathcal{F}^{-1}\left[G^{(0)}_{ij}({\bf k})\circ \mathcal{F}[\Delta \rho({\bf x}) u_j({\bf x})]\right]  = \mathcal{F}^{-1}\left[H_{iJ}^{(0)}({\bf k}) \circ M_J({\bf k})\right],    
 \end{split}
\end{equation}
where $\mathcal{F}$ and $\mathcal{F}^{-1}$ are the 3-dimensional fast Fourier and inverse fast Fourier transforms, respectively, ${\bf x}$ is a position vector in the real space and `$\circ$' denotes the elementwise multiplication operation. Although the formulation of the particle displacement has been obtained using the continuous form of FFT over the spatial domain, the discrete form of FFT is needed to perform the computational operation. The background medium should be homogeneous to apply the Fourier transform. \\[4pt]

Equation (21) can be written in the form ${\bf Au} = {\bf u}^{(0)}$. Here, $ {\bf A}$ can be regarded as an operator acting on the elastic displacement component, and $ {\bf u}^{(0)}$ is the displacement field component in the homogeneous background medium. Instead of using a direct solver, we solve the system of equations in an iterative way using the conjugate-gradient (CG) method (Shewchuk, 1994; Nocedal and Wright, 2000). When the coefficient matrix is nonsymmetric and nonsingular, variants of CG such as the BiConjugate Gradient (BiCG), Biconjugate Gradient Stabilized (BiCGSTAB) and Conjugate Gradient Squared (CGS) methods are useful (Barrett et al., 1994). We apply the BiCGSTAB method for the numerical solution, as this method has a faster and smoother convergence than the other variants of the conjugate-gradient method (Van der Vorst, 1992).
The relative residual error (e) at each iteration step can be given as:
\begin{equation}
  e = \frac{\lVert {\bf u}^{(0)} - {\bf Au}\rVert}{\lVert {\bf u}^{(0)}\rVert}.  
\end{equation}
We test the convergence of our scheme by plotting the relative residual error with the iteration numbers.

\subsection{Absorbing boundary condition}
One of the challenges in simulating the seismic wavefield in any model is to avoid reflections at the boundaries. In the finite difference method, an artificial absorbing boundary can be implemented by either using the damping boundary condition or perfectly matched layers (Lei Li et al., 2021). Alles and van Dongen (2011) developed perfectly matched layers (PMLs) for frequency-domain integral equation acoustic scattering problems. A strong attenuation of scatter pressure fields is achieved in layers with a thickness of less than a wavelength by using a plane-wave function that has different solutions within and outside the PML domain. Osnabrugge et al. (2016) used a polynomial-based expression for the wavefield inside the absorbing layer for a similar type of scattering problem and found an expression for the scattering potential of the layer in terms of the solution for the polynomial. The idea was to design a layer that has zero reflectivity for a normal incidence.\\[4pt]

Green's function approach to calculate the seismic wavefield incorporates the exponential decay of the wavefield with the distance. Most of the absorbing boundary conditions given in the differential form can be transformed to an exponential decay in the integral equation. Strictly speaking, it is difficult to completely avoid any reflections from the boundaries, however, the amplitude of the reflected wave can be significantly minimized. The contrast between the actual medium and the background medium should be zero to have no reflections at all. In other words, the actual medium wavefield is the same as the background field. For the integral equation in the particle displacement, the absorbing boundary condition is introduced by formulating the scattered field equation such that it gradually vanishes as it approaches the boundaries. \\[4pt]

Let $\Omega$ be the domain of the scattered region, $\Omega_B$ be the domain of the boundary layer region for a model and ${\bf x_B}$ be a position within $\Omega_B$. Let ${\bf x_L}$ be the position vector for a point obtained by extending the position vector ${\bf x_B}$ to the external boundary of the model. Then,
\begin{equation}
     l = \lVert {\bf x_L} - {\bf x_B}\rVert.
 \end{equation}
We can write the integral equation for the displacement as:
\begin{equation}
  u_{i}({\bf x}) = u_{i}^{(0)}({\bf x}) + \Delta u_{i}^{(\Omega)}({\bf x}).   
 \end{equation}

Then, the absorbing boundary condition can be applied through the following numerical expression:
 \begin{equation}
 \begin{split}
  \Delta u_{i}^{(\Omega_B)}({\bf x}) & = \int_{\Omega_B} d{\bf x_B} H_{ijk}^{(0)}({\bf x},{\bf x_B})\Delta c_{jklm}({\bf x_B})\epsilon _{lm}({\bf x_B})\left[1-e^{-\frac{l^2}{N_i h^2}}\right]  \\
  &  + \omega^2 \int_{\Omega_B} d{\bf x_B} G^{(0)}_{ij}({\bf x},{\bf x_B})\Delta\rho ({\bf x_B}) u_{j}({\bf x_B})\left[1-e^{-\frac{l^2}{N_i h^2}}\right].
 \end{split}
\end{equation}

Here, $h$ is the grid spacing, and $N_i$ is the number of grid points in the $i$ direction.

\section{NUMERICAL RESULTS AND DISCUSSION}

We perform all the numerical simulations for a 3D Cartesian mesh. The grid is uniformly spaced in each direction. The volume of each cell is the product of the grid spacing in the X, Y and Z-directions. Each node characterizes a unique stiffness matrix and a density value. The physical properties do not vary laterally in the Y-direction. We generally show snapshots of the models on the XZ-plane. The microseismic source is positioned at the centre of the model unless otherwise stated. To obtain an accurate scattered field, the grid size is chosen as one-fourth or smaller than the smallest wavelength of the S-waves in the actual medium.

\subsection{Model 1: Homogeneous VTI model}
The homogeneous model represents the transversely isotropic model with a vertical axis of symmetry (VTI). The density value is 2500 kg/m$^3$ (constant within the model), and the values for the stiffness coefficients for the VTI medium (Jakobsen and Johansen, 2000: Table 3, sample at a depth of 3492 m) in units of GPa are
\[ c_{IJ} = 
\left[
\begin{array}{cccccc}
34.0 & 10.6 & 6.9 & 0 & 0 & 0\\
 & 34.0 & 6.9 & 0 & 0 & 0\\
 &  & 26.5 & 0 & 0 & 0\\
 &  &  & 10.4 & 0 & 0\\
 &  &  &  & 10.4 & 0\\
 &  &  &  &  & 11.7
 \end{array}
\right].
    \]
This VTI medium is the actual medium in the integral equation approach, and it is decomposed into an isotropic background medium and a contrast medium. For the isotropic background medium, the stiffness coefficients $c_{11}$ and $c_{44}$ are 26.5 GPa and 10.4 GPa, respectively. The numerical values of the density at every grid point are the same in the background medium and the actual medium. The dimension of the model is 640 m in each direction, and the grid size is 10 m.

\subsubsection{Displacement wavefields due to a moment tensor source}

We want to demonstrate the microseismic wavefield in isotropic background and actual VTI media. The absolute value of the elastic wavefield normalized with respect to the background medium wavefield on a decibel scale is given as
\begin{equation}
 u_i^{\mbox{norm}}= 10\mbox{log}_{10}\left[\left| \frac{u_i}{\lVert u^{(0)}_i \rVert} \right|\right].
\end{equation}
\\
The wavefront for a vector point source in an isotropic media is spherical. However, for the moment tensor source, the wavefront is lobe shaped. In the time-domain snapshot (Fig. 1), the wavefront has a different polarity in opposite directions, which means that the pattern of compression and rarefaction advances in a different fashion around the source. Figs. 2 to 5 show the displacement field generated by a moment tensor source on the XZ-plane passing through the source at a particular frequency (20 Hz in this simulation). Fig. 6 shows the strain field generated by an explosion type of moment tensor source in the isotropic background medium. We label the strain field components in the Voigt notation.
In Fig. 7, we present the elastic wavefield in the VTI medium due to a double-couple source. The numerical expression for the source term is obtained following Grechka (2020). We set the tolerance limit for the relative residual error in the iterative method to a very low value, on the order of $10^{-10}$. The numerical scheme for this simulation converges in a few iterative steps (Fig. 8).

\subsubsection{Comparison with the finite difference time domain method}
To verify our methodology, we compare our results to the results obtained by using the finite difference method in the time domain. The order of the differential operator, the discretization technique and the fashion in which the source and its adjacent nodes are treated can significantly affect the result of the finite difference method. The higher the order of the differential operator is, the more accurate the wavefields obtained. However, the computational cost increases as more floating-point operations are needed (Sei and Symes, 1995). The finite difference solution to the elastic wave equation with the moment tensor source implementation can be obtained using the velocity-stress formulation on a staggered grid (Shi P. et al., 2018). The particle displacement can be obtained from the particle velocities by performing the integration over time. The elastic displacement field at the receiver position due to a moment tensor source, obtained by the integral equation (IE) method, is in the frequency domain. It is based on the evaluation of Green's tensor and its derivative. We need time domain data to compare this result with the wavefield of the displacement components obtained by the finite difference time domain (FDTD) method for the same model setup (Table 1). We set the values for the total time of simulation and the time interval equal to the respective values used in the FDTD method. The chosen time step satisfies the Courant--Friedrichs--Lewy (CFL) condition.\\[4pt]

First, we take a loop over positive and negative frequencies up to the Nyquist frequency and compute the wavefield corresponding to each frequency. Then, we perform the inverse Fourier transform (iFT) to generate time-domain particle displacement data. The waveforms of the elastic displacement field obtained using the integral equation method and the FDTD method at a common receiver match well (Fig. 9).

\subsection{Model 2: Heterogeneous VTI model}
The 3D model (Fig. 10) has three layers. An isotropic layer is sandwiched between two VTI layers. This model represents an underburden-reservoir-overburden rock system. The dimension of the model is 320 m in each direction, and the grid size is 2.5 m. The thickness of the top and bottom layers is 108 m. The middle layer is 104 m thick. The density value is 2500 kg/m$^3$ and is constant within the model. The values of the stiffness coefficients (Jakobsen and Johansen, 2000: Table 3) in the different layers are given in Table 2.

\subsubsection{Displacement wavefields due to a moment tensor source}

In Fig. 11, we see the displacement field generated by an explosion-type moment tensor source on the XZ-plane passing through the source at a frequency of 50 Hz. We select the high frequency to demonstrate the scattering of waves from the layer interfaces. The strain field components $E_1$ to $E_6$ are shown in Fig. 12. A significant amount of energy remains confined to the layer in which the source is present, i.e., in the middle layer. The numerical convergence of the scheme is presented in Fig. 13.

\subsubsection{Comparison with the finite difference time domain method}
We perform a comparison between the integral equation method and the finite difference time domain method to test the accuracy of the numerical scheme in heterogeneous media. The model parameters used in the comparison are given in Table 3. We obtain a satisfactory match between the synthetic seismograms obtained using these two methods (Fig. 14). The amplitude and phase information from the full waveform modelling can be used to demarcate the reflection and transmission events in the layered model. For this model, the strong amplitudes recorded at the receiver positions represent either the arrival of the direct waves or the arrival of the waves transmitted through layer interfaces, and the weak amplitudes are due to the arrival of the waves reflected from layer interfaces.

\subsection{Model 3: Hess model}
We also apply our numerical scheme on the resampled Hess/SEG salt model (Fig. 15). The grid size is 10 m. We use the 2D dataset and extend the model in the Y-direction to 1 km to make it 3D. We want to observe the effect of the lateral variations in the physical properties of the medium on the wavefield. The density values vary within the medium. The model contains a fault zone. Along with the sedimentary layers, a salt body is also present. Hence, the geology of this model is much more complex than that of Models 1 and 2.

\subsubsection{Wavefield of displacement components}
We observe the wavefield on the XZ-plane passing through the microseismic source. The strain field components in the Voigt notation, $E_1$ to $E_6$ are shown in Fig. 16. The seismic source is an explosion-type moment tensor. The X-, Y- and Z-components of the elastic displacement written here as $u_x$, $u_y$ and $u_z$, respectively, are shown in Fig. 17. The presence of the salt body and the fault can be inferred by observing the strain field components. The importance of the strain field calculation is that the strain components are needed during the inversion for the model parameters. \\[4pt]

Although a large number of iterations are needed, the numerical scheme converges for this complex model (Fig. 18). Furthermore, we tested our scheme by setting the contrast between the stiffness coefficient values at every grid point in the actual medium and the homogeneous background medium to be up to $10\% $ and found that it still converges.

\section{CONCLUDING REMARKS}
We have presented an effective volume integral equation method to model elastic wavefields due to the moment tensor source. The integral equation method for microseismic wavefield modelling can in principle be applied to general anisotropic media, although our numerical experiments involved VTI media only. When the background medium and the actual medium are significantly different, the Born approximation fails to accurately produce the scattered wavefield. The Fourier transform-based iterative solver can be applied even when the contrast between the actual medium and the background medium is high, and we can obtain a full solution of the Lippmann-Schwinger equation for elastic waves. We have demonstrated here the technique to perform seismic modelling efficiently in a matrix-free manner. By using a fast algorithm based on the discrete Fourier transform and the biconjugate gradient stabilized (Bi-CGSTAB) method, one can accelerate the iterative solution of the integral equations. A further acceleration of our fast IE method may be achieved by the use of preconditioners. \\[4pt]

The generation of synthetic seismic data for the complex heterogeneous model emphasizes the utility of the volume integral method for modelling elastic wavefields in complicated three-dimensional geological structures. Wavefield computation at multiple frequencies can be performed simultaneously, and therefore, we think that this approach is favourable for parallel computing operations. The present methodology is suitable to generate a forward model for the frequency domain full waveform inversion for moment tensor components.

\section*{ACKNOWLEDGEMENTS}
The authors acknowledge the VISTA program, The Norwegian Academy of Science and Letters and Equinor, for their financial support of this project.

\section*{DATA AVAILABILITY}
In our numerical simulation, we need the density and stiffness coefficient values within a medium. We used three models, namely, Model 1, Model 2 and Model 3. The details of Models 1 and 2 are given in the numerical results section of the paper. Model 3 is the SEG/EAGE salt (Hess) model, which can be downloaded from the SEG website link (https://software.seg.org/datasets/2D/Hess\_VTI/).

\newpage
\section*{REFERENCES}

Achenbach, J. D., 1975.
Wave propagation in elastic solids,
\textit{Elsevier science publishers B.V.}, ISBN: 07204 03251.\\[8pt]

Aki, K., and Richards, P. G., 2002.
Quantitative Seismology,
\textit{University Science Books}, Mill Valley, California. \\[8pt]

Alles, E. J. and van Dongen, K. W. A., 2011.
Perfectly matched layers for frequency-domain integral equation acoustic scattering problems,
\textit{IEEE transactions on ultrasonics, ferroelectrics, and frequency control}, vol. \textbf{58}, Issue 5.\\[8pt]

Auld, B., 1990.
Acoustic Fields and Waves in Solids,
\textit{Krieger Publishing Company}, ISBN: 0894644904, 9780894644900.\\[8pt]

Barrett, R., Berry, M., Chan, T. F., Demmel, J., Donato, J. M., Dongarra, J., Eijkhout, V., Pozo, R., Romine, C., and Vorst, H. V., 1994.
Templates for the Solution of Linear Systems: Building Blocks for Iterative Methods,
\textit{Society for Industrial and Applied Mathematics}, (http://www.siam.org/books).\\[8pt]

Červený , V. and Ravindra, R., 1971.
Theory of Seismic Head Waves,
\textit{Univ. of Toronto Press}, Toronto.\\[8pt]

Červený , V. and Hron, F., 1980.
The ray series method and dynamical ray tracing system for three-dimensional inhomogeneous media,
\textit{Bull. Seismol. Soc. Am.}, 70, 47--77.\\[8pt]

Červený , V., 1985.
Ray synthetic seismograms for complex two-and three-dimensional structures,
\textit{J. Geophys}, 58.\\[8pt]

Červený , V., 2001.
Seismic ray theory,
\textit{Cambridge University Press}.\\[8pt]

Gibson, R. L. and Ben-Menahem, A., 1991.
Elastic wave scattering by anisotropic obstacles: Application to fractured volumes,
\textit{Journal of Geophysical Research}, vol. \textbf{96}, No. B12.\\[8pt]

Graves, R. W., 1996.
Simulating seismic wave propagation in 3D elastic media using staggered-grid finite differences,
\textit{Bulletin of the Seismological Society of America}, vol. \textbf{86}, No. 4, pp. 1091-1106.\\[8pt]

Grechka, V., 2020.
Moment tensors of double-couple microseismic sources in anisotropic formations,
\textit{Geophysics}, vol. \textbf{85}, No. 1, KS1-KS11.\\[8pt]

Hudson, J. A. and Heritage, J. R., 1982.
The use of the Born approximation in seismic scattering problems,
\textit{Geophys. J. R. Astron. Soc.}  \textbf{66}, 221--240.\\[8pt]

Jakobsen, M. and Johansen, T. A. , 2000.
Anisotropic approximations for mudrocks: A seismic laboratory study,
\textit{Geophysics}, vol. \textbf{65} No. 6, P. 1711--1725.\\[8pt]

Jakobsen, M. and Ursin, B. , 2015.
Full waveform inversion in the frequency domain using direct iterative T-matrix methods,
\textit{Journal of Geophysics and Engineering}, \textbf{12(3)}, 400-418.\\[8pt]

Jakobsen, M. and Tveit, S. , 2018.
Distorted Born iterative T-matrix method for inversion of CSEM data in anisotropic media,
\textit{Geophys. J. Int.}, \textbf{214}, 1524--1537.\\[8pt]

Jakobsen, M., Psencík, I., Iversen, E. and Ursin, B. , 2020.
Transition operator approach to seismic full-waveform inversion in arbitrary anisotropic elastic media,
\textit{Commun. Comput. Phys.}, Vol. \textbf{28}, No. 1, pp. 297-327.\\[8pt]

Jost, M. and Herrmann, R., 1989.
A student's guide to review of moment tensors,
\textit{Seismol Res Letter}, \textbf{60(2)}, 37--57.\\[8pt]

Lei, L., Tan, J., Zhang, D., Malkoti, A., Abakumov, I. and Xie, Y., 2021.
FDwave3D: a MATLAB solver for the 3D anisotropic wave equation
using the finite-difference method,
\textit{Computational Geosciences}, \textbf{25}: 1565--1578.\\[8pt]

Love, A. E. H., 1944.
The mathematical theory of elasticity,
\textit{4th ed. New York, Dover Publications, Inc.}\\[8pt]

Madariaga, R., 2015.
Seismic source theory,
\textit{Elsevier B.V.}\\[8pt]

Malovichko, M., Khokhlov, N., Yavich, Zhdanov, M., 2018.
Acoustic 3D modeling by the method of integral equations,
\textit{Computers and Geosciences}, \textbf{111}, 223--234.\\[8pt]

Miles, J. W., 1960.
Scattering of elastic waves by small inhomogeneities,
\textit{Geophys.}, \textbf{25}, 642-648.\\[8pt]

Monchiet, V. and Bonnet, G., 2012.
A polarization based FFT iterative scheme for computing the effective properties of elastic composites with arbitrary contrast,
\textit{International Journal for Numerical Methods in Engineering}, Wiley, 2012, \textbf{89}(11), pp.1419-1436.\\[8pt]

Nocedal, J. and Wright, S. J., 2000.
Numerical Optimization,
\textit{Springer Science$+$Business Media, LLC.}, ISBN-10: 0-387-30303-0. \\[8pt]

Osnabrugge G., Leedumrongwatthanakun S., Vellekoop I.M., 2016.
A convergent Born series for solving the inhomogeneous Helmholtz equation in arbitrarily large media,
\textit{Journal of Computational Physics}, vol. \textbf{322}.\\[8pt]

Pujol, J., 2003.
Elastic wave propagation and generation in seismology,
\textit{Cambridge university press}.\\[8pt]

Rice, J. R., 1998.
Notes on elastodynamics, Green's function, and response to transformation strain and crack or fault sources,
\textit{course: Earth and Planetary Sciences 263}, Harvard University.\\[8pt]

Sei, A. and Symes, W., 1995.
Dispersion analysis of numerical wave propagation and its computational consequences,
\textit{Journal of Scientific Computing}, \textbf{Vol 10}, No. 1.\\[8pt]

Shewchuk, J. R., 1994.
An introduction to the Conjugate Gradient method without the agonizing pain,
\textit{School of Computer Science}, Carnegie Mellon University Pittsburgh.\\[8pt]

Shi, P., Angus, D., Nowacki, A., Yuan, S., Wang, Y., 2018.
Microseismic full waveform modeling in anisotropic media with moment tensor implementation,
\textit{Surveys in Geophysics}.\\[8pt]

Snieder, R., 2002.
General theory of elastic wave scattering., Scattering and Inverse Scattering in Pure and Applied Science, Pages 528-542, \textit{Elsevier Ltd.}\\[8pt]

Stein, S. and Wysession, M., 2003.
An introduction to seismology, earthquakes, and earth structure,
\textit{Blackwell publishing ltd}, ISBN 0-86542-078-5.\\[8pt]

Tong, M. S. and Chew, W. C., 2009.
Multilevel fast multipole algorithm for elastic wave scattering by large three-dimensional objects,
\textit{Journal of Computational Physics}, vol. \textbf{228}, Issue 3.\\[8pt]

Touhei, T., 2011.
A fast volume integral equation method for elastic wave propagation in a half space,
\textit{International Journal of Solids and Structures}, vol. \textbf{48}, Issues 22-23.\\[8pt]

Lecture notes on Fourier transform in N dimensions,
\textit{Departments of Radiology and Medical Physics}, University of Wisconsin-Madison.\\[8pt]

Van der Vorst, H. A., 1992.
Bi-CGSTAB: A fast and smoothly converging variant of Bi-CG for the solution of nonsymmetric linear systems,
\textit{SIAM J.Sci. Stat Comput.}, \textbf{B(2)}, 631-644.\\[8pt]

Virieux, J., 1986.
P-Sv wave propagation in heterogeneous media: Velocity-stress finite-difference method,
\textit{Geophysics}, \textbf{51(4)}, 889--901.\\[8pt]

Wu, R. S. and Ben-Menahem, A., 1985.
The elastodynamic near field,
\textit{Geophys. J. R. Astron. Soc.}, \textbf{81}, 609--622.\\[8pt]

Xiang, K., Eikrem, K. S., Jakobsen, M. and Nævdal, G., 2021.
Homotopy scattering series for seismic forward modelling with variable density and velocity,
\textit{Geophysical Prospecting}, vol. \textbf{70(1)}, 3-18.\\[8pt]

\newpage
\renewcommand{\theequation}{A.\arabic{equation}}
\setcounter{equation}{0}
\subsection*{APPENDIX A: ELASTIC GREEN'S FUNCTION AND ITS SPATIAL DERIVATIVE}

Many authors have presented the mathematical expression for the elastic Green's function. The expressions can be nonidentical, as they are derived in light of different underlying physics. As an example, the Green's function given by Červený and Ravindra (1971), Červený and Hron (1980) considered the ray-geometric propagation of body waves in elastic media. In the far field, terms corresponding to P- and S-wave contributions can simply be added to formulate Green's tensor. However, in the near field, it is not possible to separately write a Green's tensor expression for the P and S waves. This can be attributed to the strong coupling effects between the P and S waves in the near field. As a high-frequency approximation, a fundamental limitation of the ray theoretical Green's tensor is the inability to model caustics or shadow zones (Červený, 1985).\\

The unit polarisation vectors for S-waves can be expressed in terms of the unit polarisation vector for the P wave. Since the direction of polarization for the P wave is radial and outwards from the source towards the receiver, the unit vector can easily be determined from the distance r between the source position ${\bf x_s}$ and the receiver position ${\bf x_r}$ in the 3D space.
\begin{equation}
  {\bf \hat{r}}  = ({\bf x_r - x_s})/r = {\bf x}/r.  
\end{equation}
The near field and far field regions can be demarcated by the relationship between the wavelength of a P-wave ($\lambda_P$) of a circular frequency ($\omega$) and the distance r between the source and the receiver. 
\begin{equation}
\lambda_P  = 2\pi \alpha /\omega.    
\end{equation}
The far field is defined as the region of space that is far away from the source compared to the wavelength (Snieder, 2002) and is given by the condition,
\begin{equation}
 \frac{2\pi r}{\lambda_P} >> 1.   
\end{equation}
For near-field radiation, Wu and Ben-Menahem (1985) derived an expression for Green's function considering the intricate relationship between the P and S waves. Green's tensor in the near field is given as
\begin{equation}
    \boldsymbol{G}^{NF} = \frac{-1}{8\pi\rho r\beta^2}\left[1- \frac{\beta^2}{\alpha^2}\right]\left[{\bf I} - \frac{3{\bf xx^T}}{r^2}\right]. 
\end{equation}
Green's tensor in the far field is given by the following equation:
\begin{equation}
    \boldsymbol{G}^{FF} = \frac{1}{4\pi\rho r\alpha^2}e^{i\omega r /\alpha}\left[\frac{{\bf xx^T}}{r^2}\right] + \frac{1}{4\pi\rho r\beta^2}e^{i\omega r /\beta}\left[ {\bf I} - \frac{{\bf xx^T}}{r^2}\right].
\end{equation}

Here, $\alpha$ and $\beta$ are the P-wave velocity and the S-wave velocity, respectively. $\rho$ is the density of the medium. ${\bf x}$ is the position vector of the receiver point with respect to the source point. The first term of Green's tensor in the far field is the far field radiation due to the P-waves, and the second term is the far field radiation due to the S-waves. The first-order spatial derivative of this Green's function can be determined using the chain rule of differentiation. \\[2pt]
We follow the same strategy as used in Jakobsen et al., 2020 and concentrate first on the near-field part of Green's function. The near-field term (Equation A.4) can be rewritten as
the following form:
\begin{equation}
    \boldsymbol{G^{NF}_{kl}} = -\frac{CE_{kl}(r)}{U(r)},
\end{equation}
where \[ C = \frac{1}{8\pi\rho \beta^2}\left(1- \frac{\beta^2}{\alpha^2}\right)\]
\[ E_{kl}(r) = {\bf I} - \frac{3x_k x_l}{r^2}\]
\[U(r) = r\]
\[U_{,i} = x_i/r\]
\[E_{kl,i} = \frac{3(2x_k x_l x_i - (\delta_{ik}x_l + \delta_{il}x_k) r^2)}{r^4}\]
\[D_i = x_i\]  
\[F_{kl} = x_k x_l\]
where $\delta_{ij}$ is Kronecker's delta function.
\begin{equation}
    \boldsymbol{G^{NF}_{kl,i}} = \frac{C}{U^3}\left[D_i E_{kl} - \frac{3}{U^2}\left(2F_{kl} D_i - (\delta_{ik}D_l + \delta_{il}D_k) U^2\right)\right].
\end{equation}
Similarly, we focus on the far field terms for the P and S waves.
\begin{equation}
    \boldsymbol{G^{P}_{kl}} = \frac{X(r)Y_{kl}(r)}{U(r)},
\end{equation}
where \[ X(r) = \frac{1}{4\pi\rho \alpha^2}e^{i\omega r /\alpha}\]
\[Y_{kl}(r) = x_k x_l /r^2\]
\[X_{,i} = \frac{i\omega X(r) x_i}{\alpha r}\]
\[Y_{kl,i} = \frac{r^2(\delta_{ik}x_l + \delta_{il}x_k) - 2x_k x_l x_i}{r^4}\]
\begin{equation}
    \boldsymbol{G^{P}_{kl,i}} = \frac{(X_{,i}Y_{kl} + XY_{kl,i})U - XY_{kl}U_{,i}}{U^2}.
\end{equation}

\begin{equation}
    \boldsymbol{G^{S}_{kl}} = \frac{V(r)W_{kl}(r)}{U(r)},
\end{equation}
where \[ V(r) = \frac{1}{4\pi\rho \beta^2}e^{i\omega r /\beta}\]
\[W_{kl}(r) = \delta _{kl} - \frac{x_k x_l}{r^2}\]
\[V_{,i} = \frac{i\omega V(r) x_i}{\beta r}\]
\[W_{kl,i} = \frac{-r^2(\delta_{ik}x_l + \delta_{il}x_k) + 2x_k x_l x_i}{r^4}\]
\begin{equation}
    \boldsymbol{G^{S}_{kl,i}} = \frac{(V_{,i}W_{kl} + VW_{kl,i})U - VW_{kl}U_{,i}}{U^2}.
\end{equation}

\newpage

\begin{table*}
\begin{minipage}{115mm}
\caption{Numerical values used in the comparison between the IE and FDTD methods for the homogeneous VTI model.}
\label{anymode}
\begin{tabular}{@{}llllll}
\hline
\hline
Modeling parameters & Values \\
\hline
Number of grid points in each direction     & Nx = Ny = Nz = 192  \\[2pt]           
Grid spacing in each direction     & dx = dy = dz = 2.5 m    \\[2pt]   
Source type  &  Moment tensor source, $M_{12} = M_{21} = 1/\sqrt{2}$  \\[2pt]
Source location   & (300, 300, 180) m    \\[2pt]   
location of downhole array of 15 receivers   & (150, 150, 265) m   \\[2pt]   
spacing between two consecutive receivers   & 10 m   \\[2pt]   
Wavelet & Ricker with peak frequency = 60 Hz      \\[2pt]
Time step &  0.0003 s  \\[2pt]
Total time of simulation &  0.24 s  \\[2pt]
\hline
\hline
\end{tabular}
\end{minipage}
\end{table*}

\begin{table*}
\begin{minipage}{115mm}
\caption{Elastic stiffnesses (in units of GPa) for Model 2 - heterogeneous VTI model.}
\label{anymode}
\begin{tabular}{@{}llllll}
\hline
\hline
Stiffness coefficients & Top VTI layer & Isotropic layer & Bottom VTI layer \\
\hline
$c_{11}$ & 34.0 & 29.05 & 33.8  \\[2pt] 
$c_{33}$ & 26.5 & 29.05 & 21.9  \\[2pt]           
$c_{55}$ & 10.4 & 10.0  &  6.0 \\[2pt] 
$c_{66}$ & 11.7 & 10.0  & 12.0 \\[2pt] 
$c_{13}$ &  6.9  & 9.05  &  8.0 \\[2pt] 
\hline
\hline
\end{tabular}
\end{minipage}
\end{table*}
\par

\begin{table*}
\begin{minipage}{115mm}
\caption{Numerical values used in the comparison between IE and FDTD methods for the heterogeneous VTI model.}
\label{anymode}
\begin{tabular}{@{}llllll}
\hline
\hline
Modeling parameters & Values \\
\hline 
Number of grid points in each direction     & Nx = Ny = Nz = 128  \\[2pt]           
Grid spacing in each direction     & dx = dy = dz = 2.5 m    \\[2pt]   
Source type  &  Moment tensor source, $M_{13} = M_{31} = 1/\sqrt{2}$  \\[2pt]
Source location   & (250, 250, 160) m    \\[2pt]   
location of downhole array of 15 receivers   & (100, 100, 120) m    \\[2pt]   
spacing between two consecutive receivers   & 10 m   \\[2pt]
Wavelet & Ricker with peak frequency = 60 Hz      \\[2pt]
Time step &  0.0003 s  \\[2pt]
Total time of simulation &  0.24 s  \\[2pt]
\hline
\hline
\end{tabular}
\end{minipage}
\end{table*}

\begin{figure*}
\centering
\begin{minipage}{0.45\textwidth}
    \includegraphics[width=2.5in]{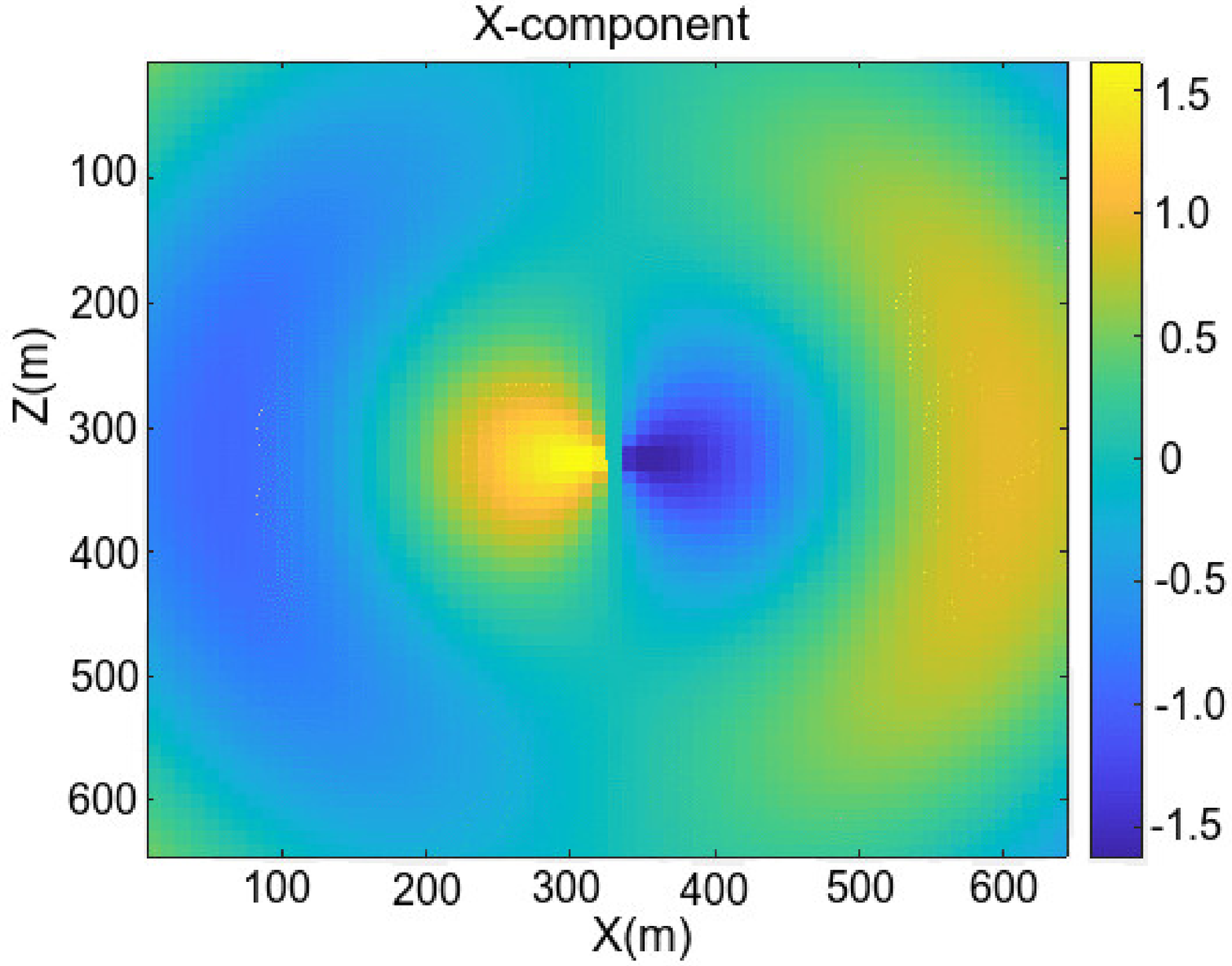}
\end{minipage}
\begin{minipage}{0.45\textwidth}
    \includegraphics[width=2.5in]{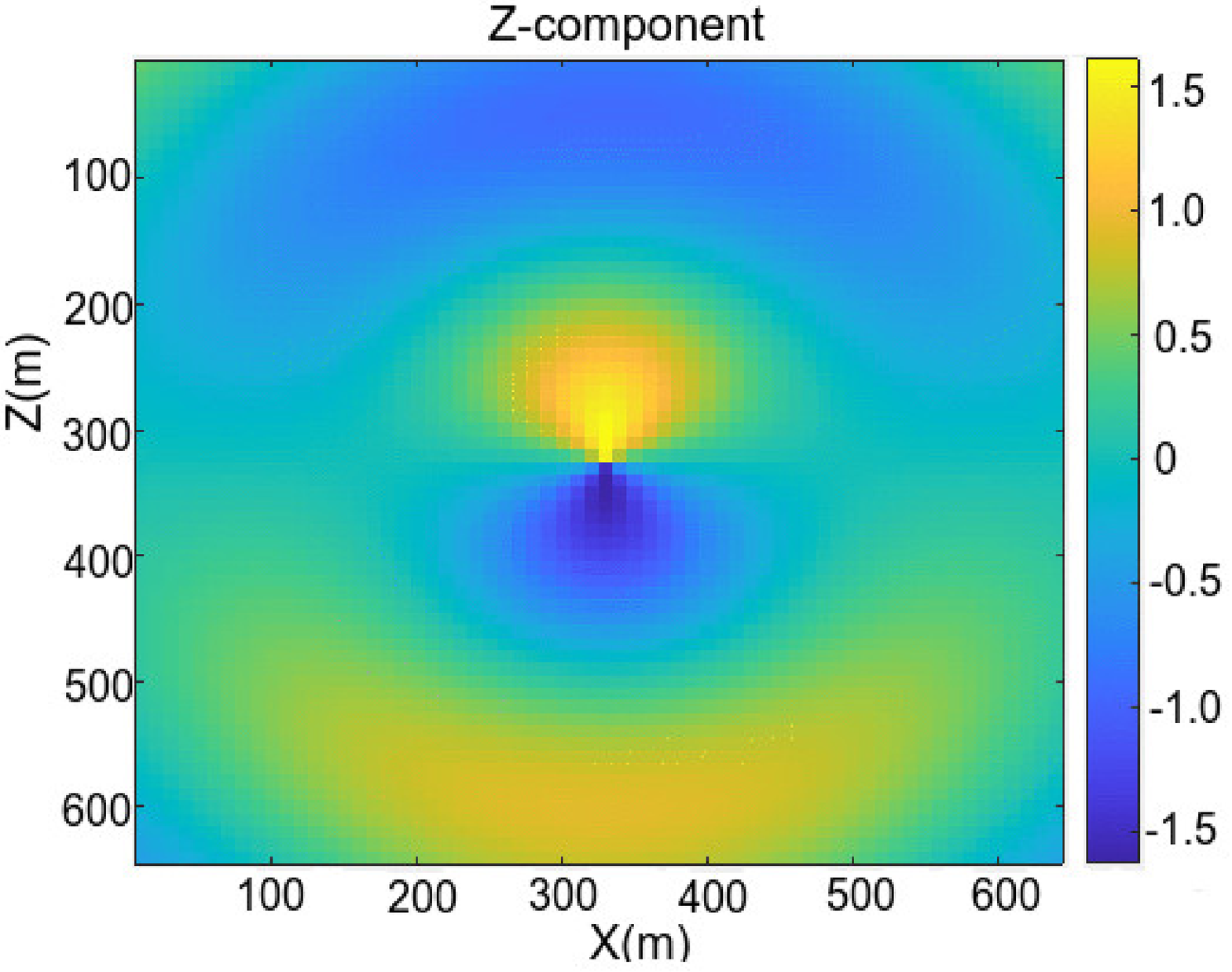}
\end{minipage}
\caption{Snapshot of the elastic displacement field in the isotropic background medium for the homogeneous model at time t = 0.12 s, when $M_{11} = M_{33} = \frac{1}{\sqrt{2}}$ and the other moment tensor components are zero.}
\end{figure*}

\begin{figure*}
\centering
\begin{minipage}{0.45\textwidth}
    \includegraphics[width=2.5in]{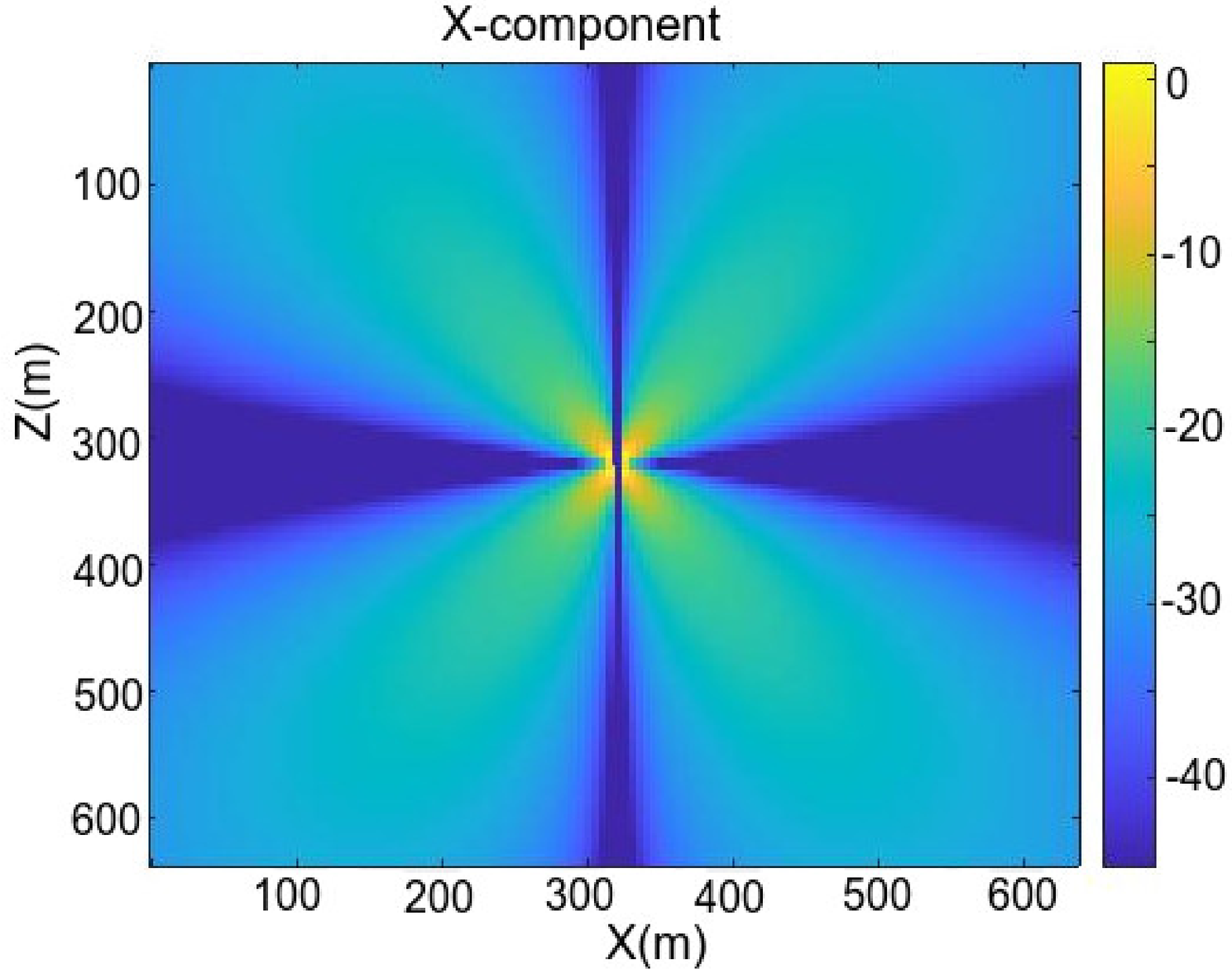}
\end{minipage}
\begin{minipage}{0.45\textwidth}
    \includegraphics[width=2.5in]{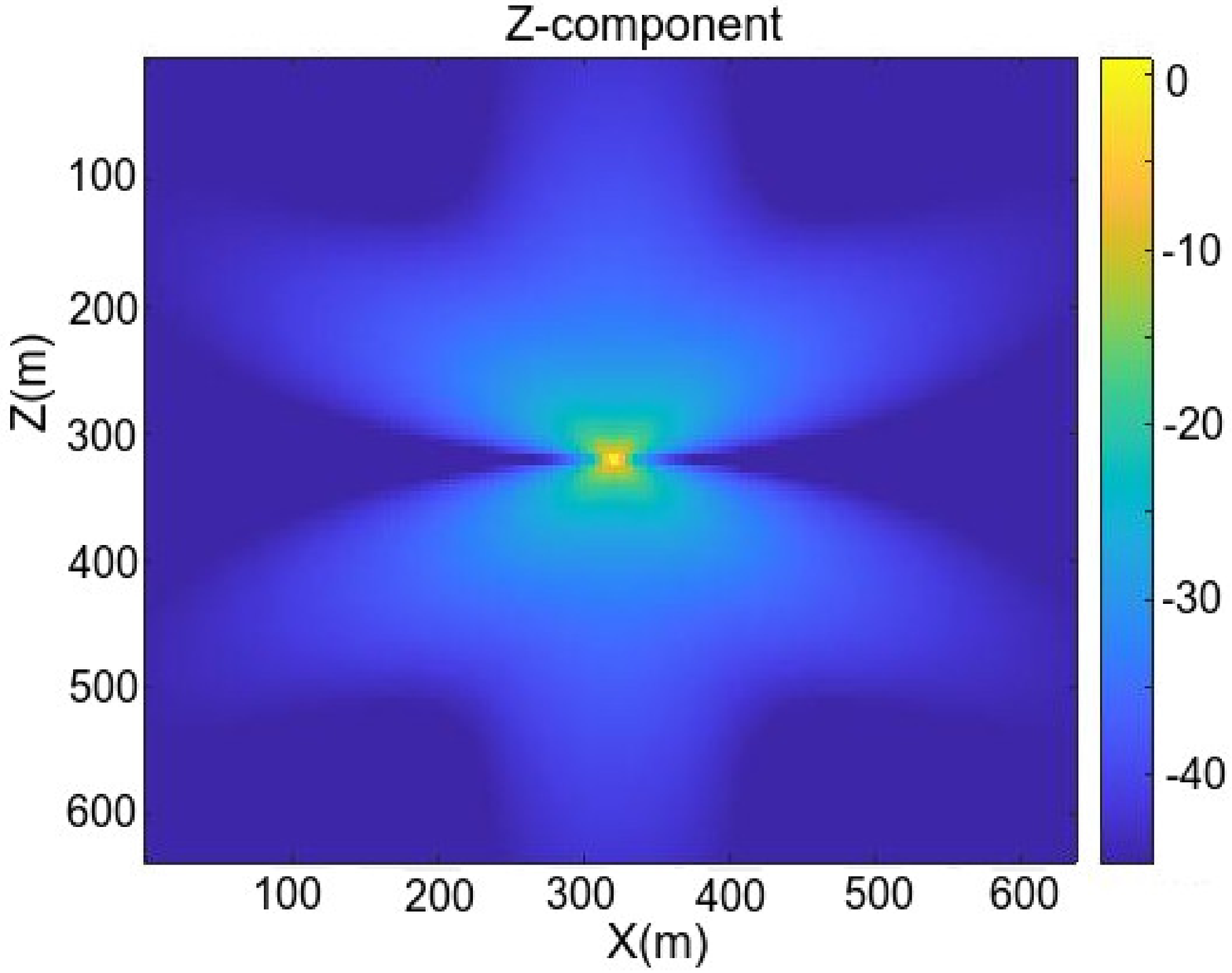}
\end{minipage}
\caption{Displacement field at 20 Hz in the isotropic background medium for the homogeneous model when $M_{33}= \frac{1}{\sqrt{2}}$ and the other moment tensor components are zero.}
\end{figure*}

\begin{figure*}
\centering
\begin{minipage}{0.45\textwidth}
    \includegraphics[width=2.5in]{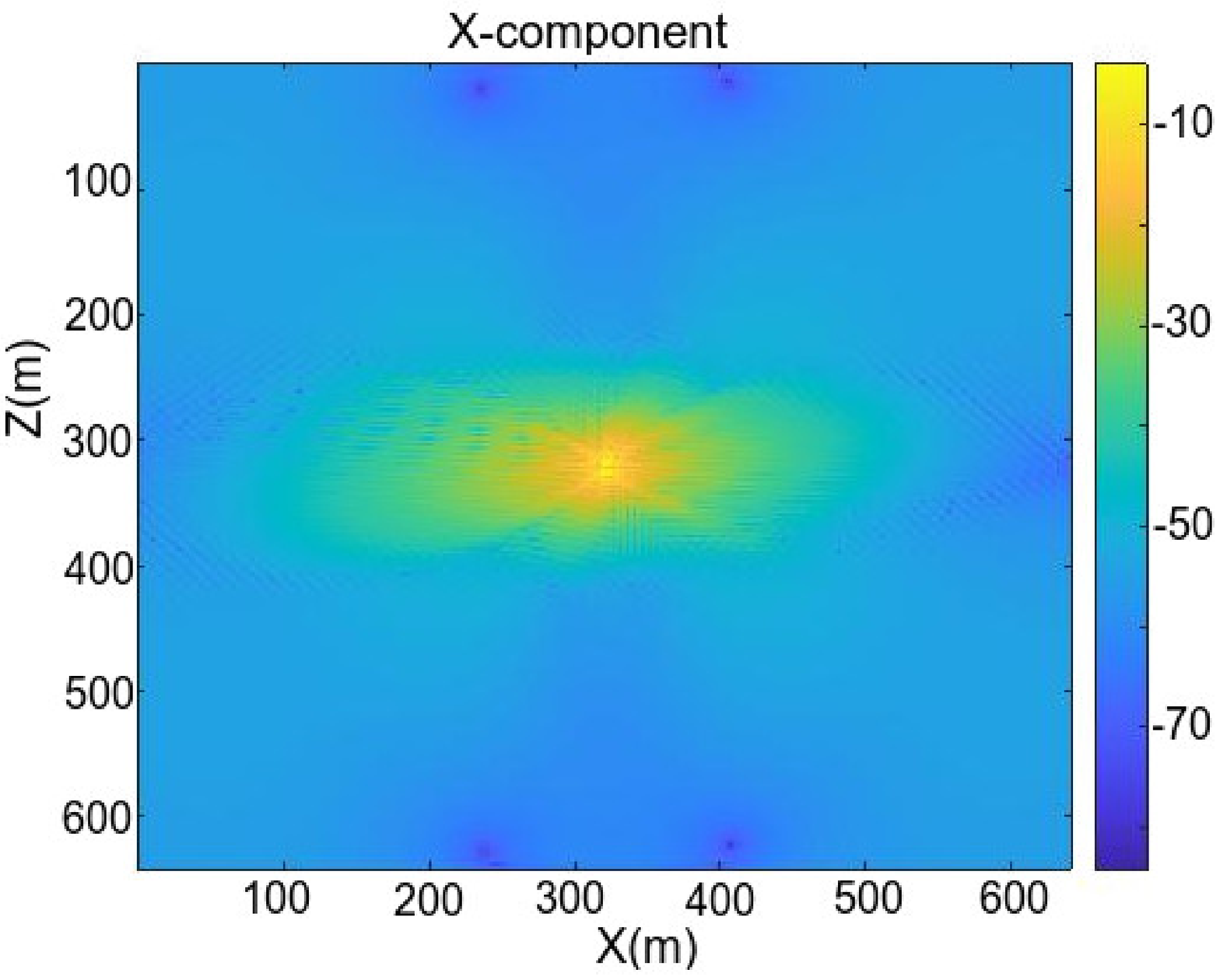}
\end{minipage}
\begin{minipage}{0.45\textwidth}
    \includegraphics[width=2.5in]{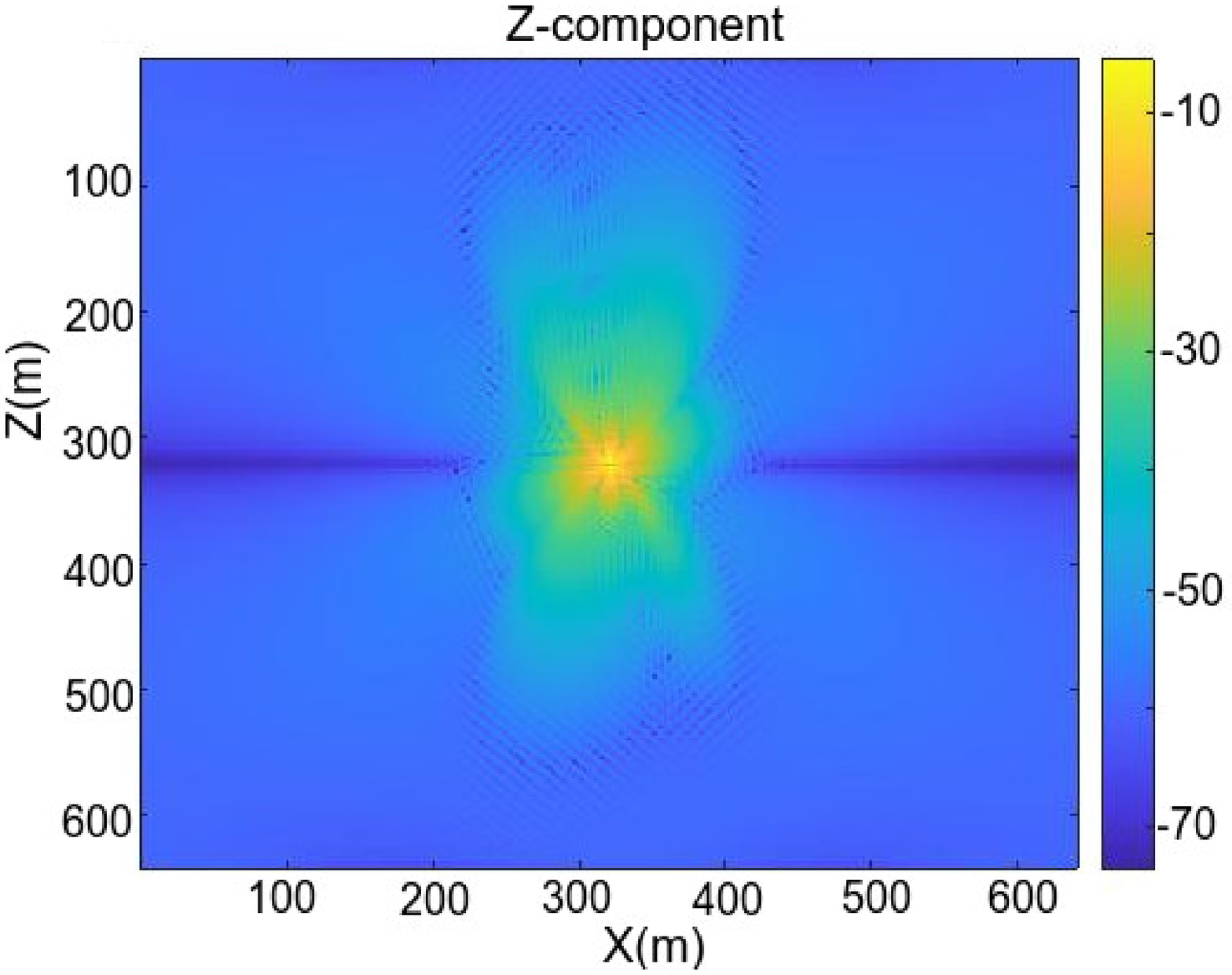}
\end{minipage}
\caption{Displacement field at 20 Hz in the homogeneous VTI medium when $M_{33}= \frac{1}{\sqrt{2}}$ and the other moment tensor components are zero.}
\end{figure*}

\begin{figure*}
\centering
\begin{minipage}{0.45\textwidth}
    \includegraphics[width=2.5in]{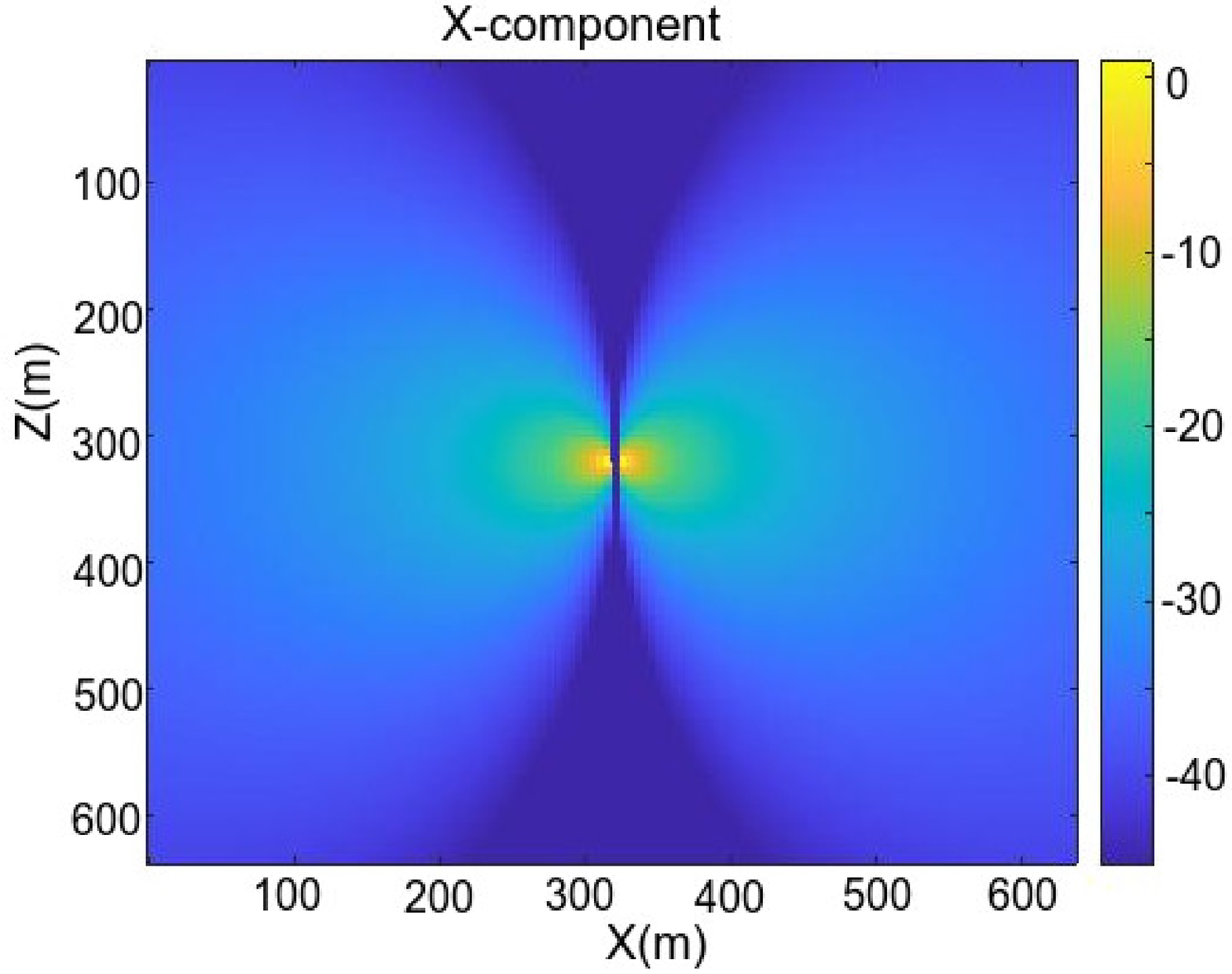}
\end{minipage}
\begin{minipage}{0.45\textwidth}
    \includegraphics[width=2.5in]{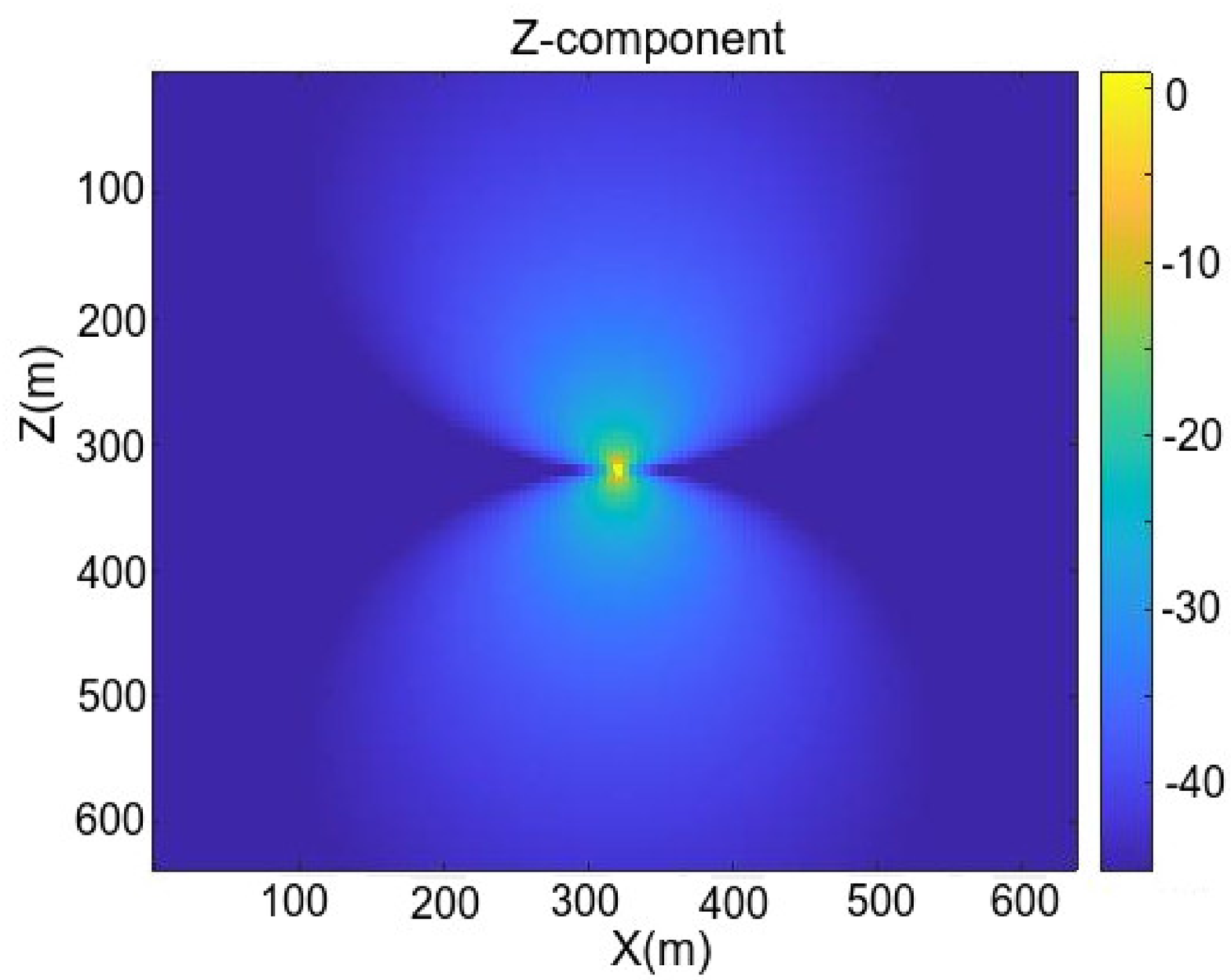}
\end{minipage}
\caption{Displacement field at 20 Hz in the isotropic background medium for the homogeneous model when $M_{11} = M_{33} = \frac{1}{\sqrt{2}}$ and the other moment tensor components are zero.}
\end{figure*}

\begin{figure*}
\centering
\begin{minipage}{0.45\textwidth}
    \includegraphics[width=2.5in]{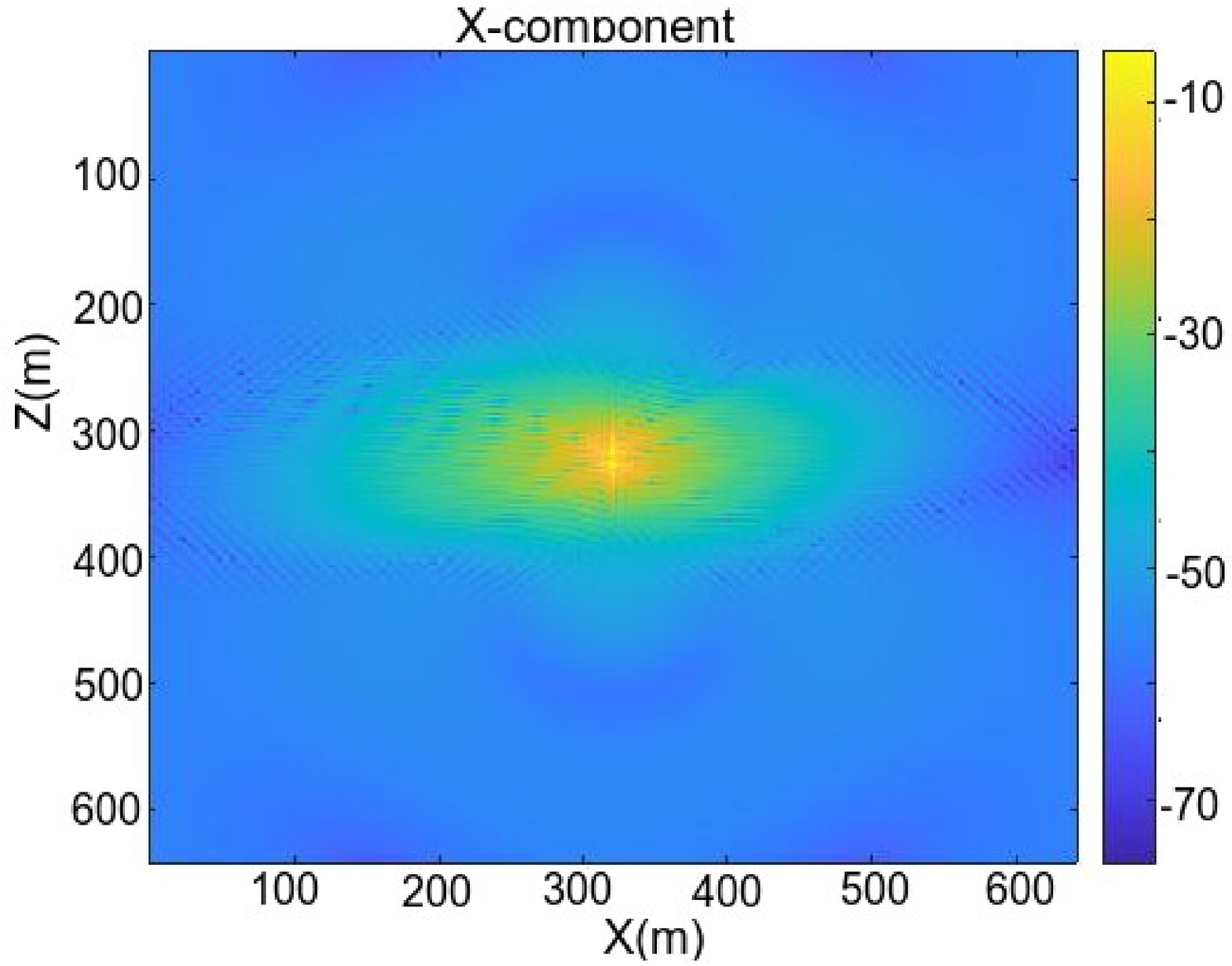}
\end{minipage}
\begin{minipage}{0.45\textwidth}
    \includegraphics[width=2.5in]{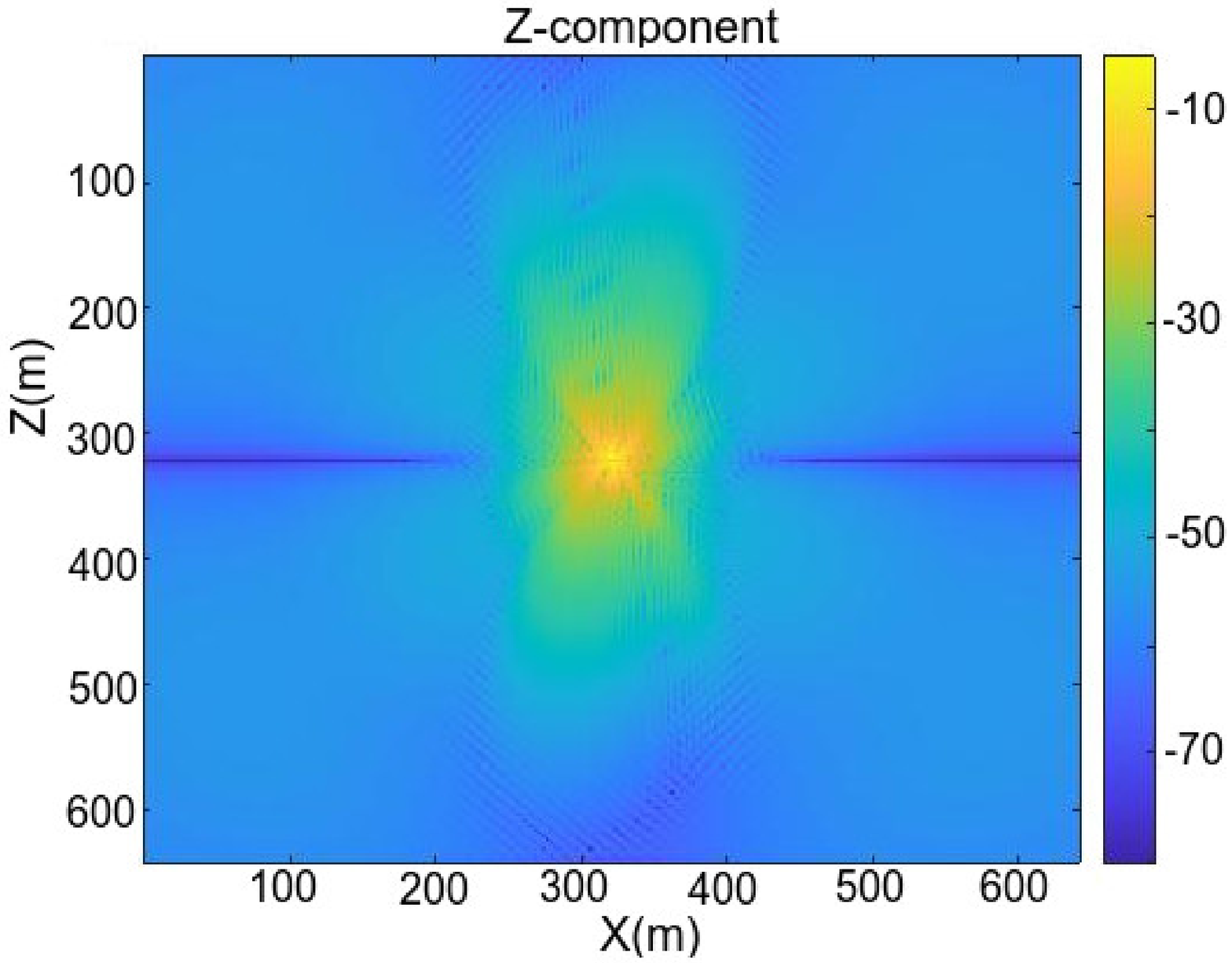}
\end{minipage}
\caption{Displacement field at 20 Hz in the homogeneous VTI medium when $M_{11} = M_{33} = \frac{1}{\sqrt{2}}$ and the other moment tensor components are zero.}
\end{figure*}

\begin{figure*}
    \centering
    \includegraphics[width=1\textwidth]{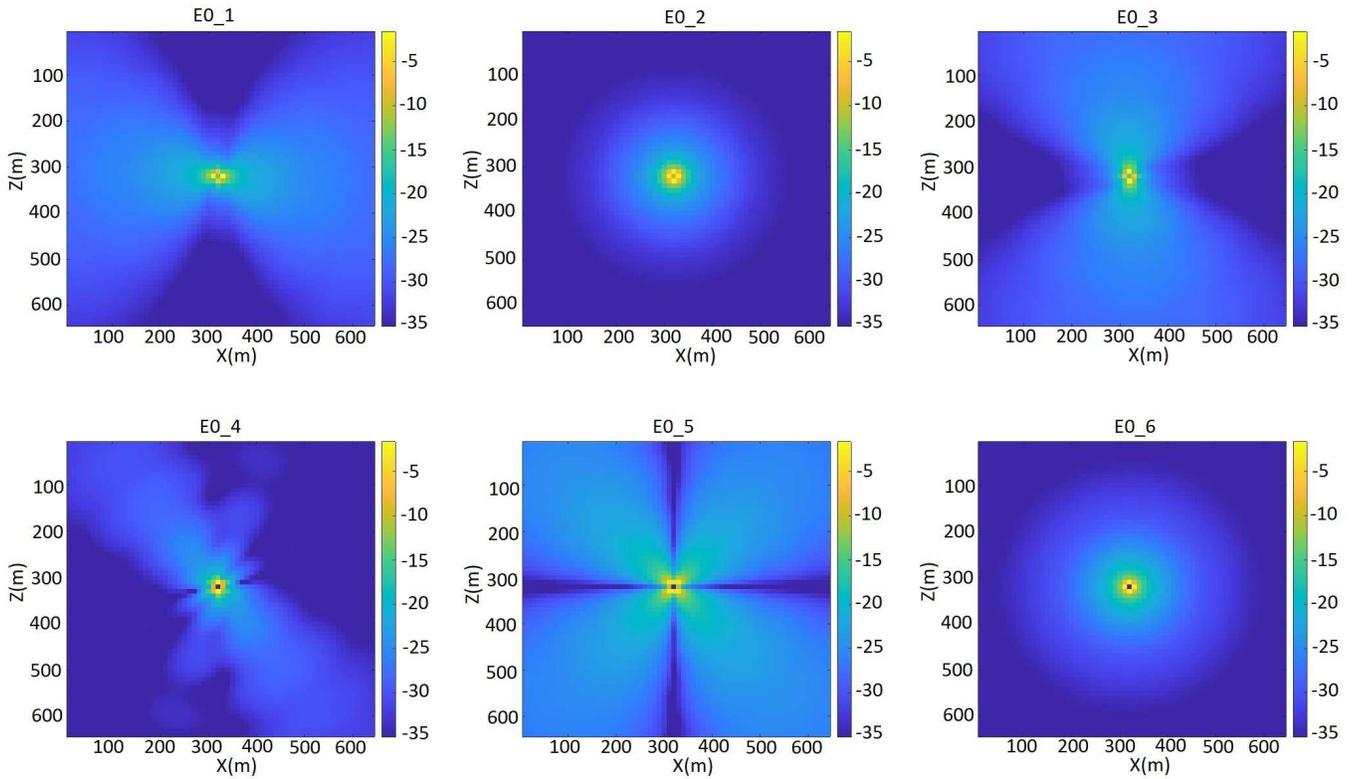}
    \caption{Strain field at 10 Hz in the homogeneous isotropic background medium due to an explosion-type moment tensor source.}
\end{figure*}

\begin{figure*}
\centering
\begin{minipage}{0.5\textwidth}
    \includegraphics[width=4.0in]{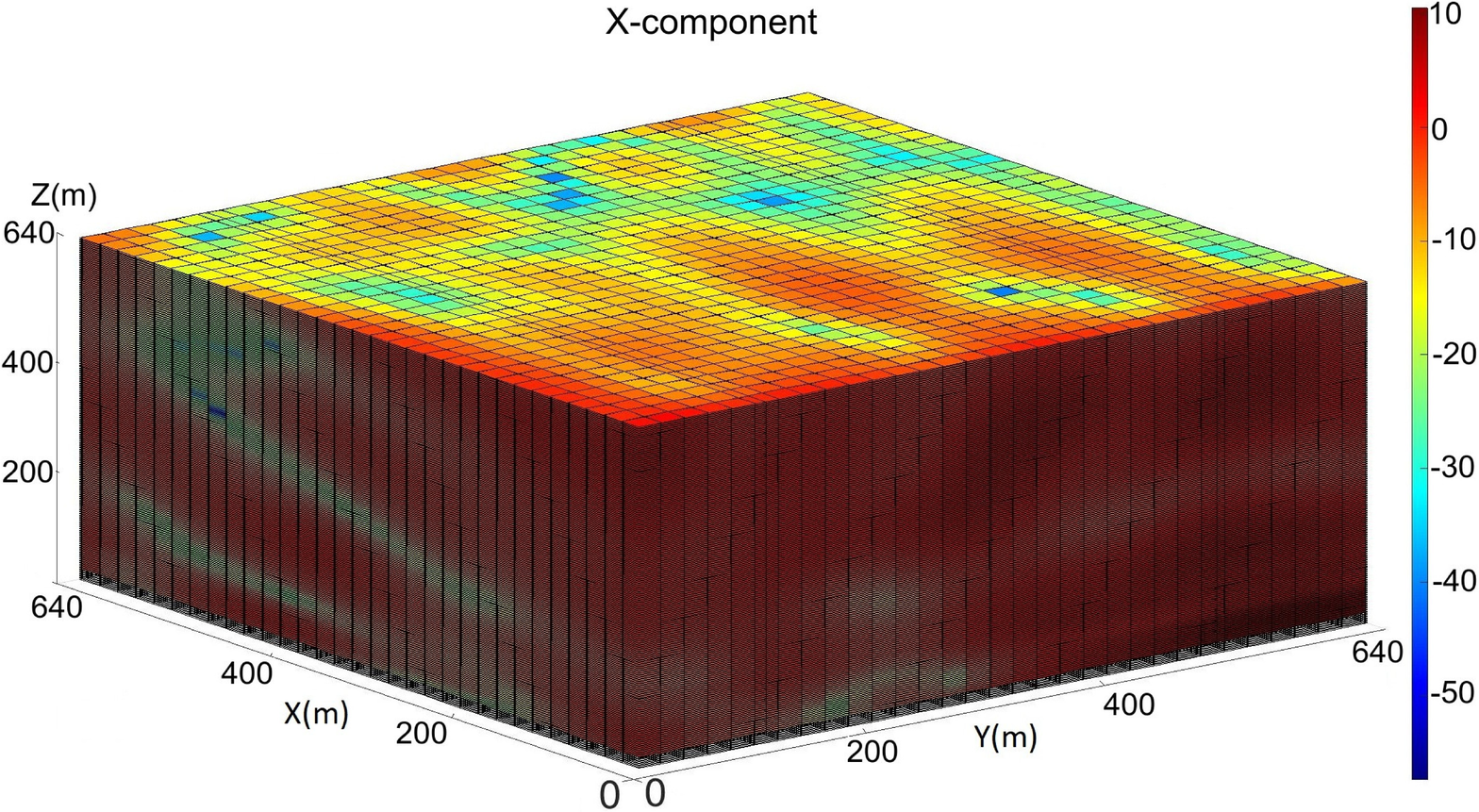}
\end{minipage}
\begin{minipage}{0.5\textwidth}
    \includegraphics[width=4.0in]{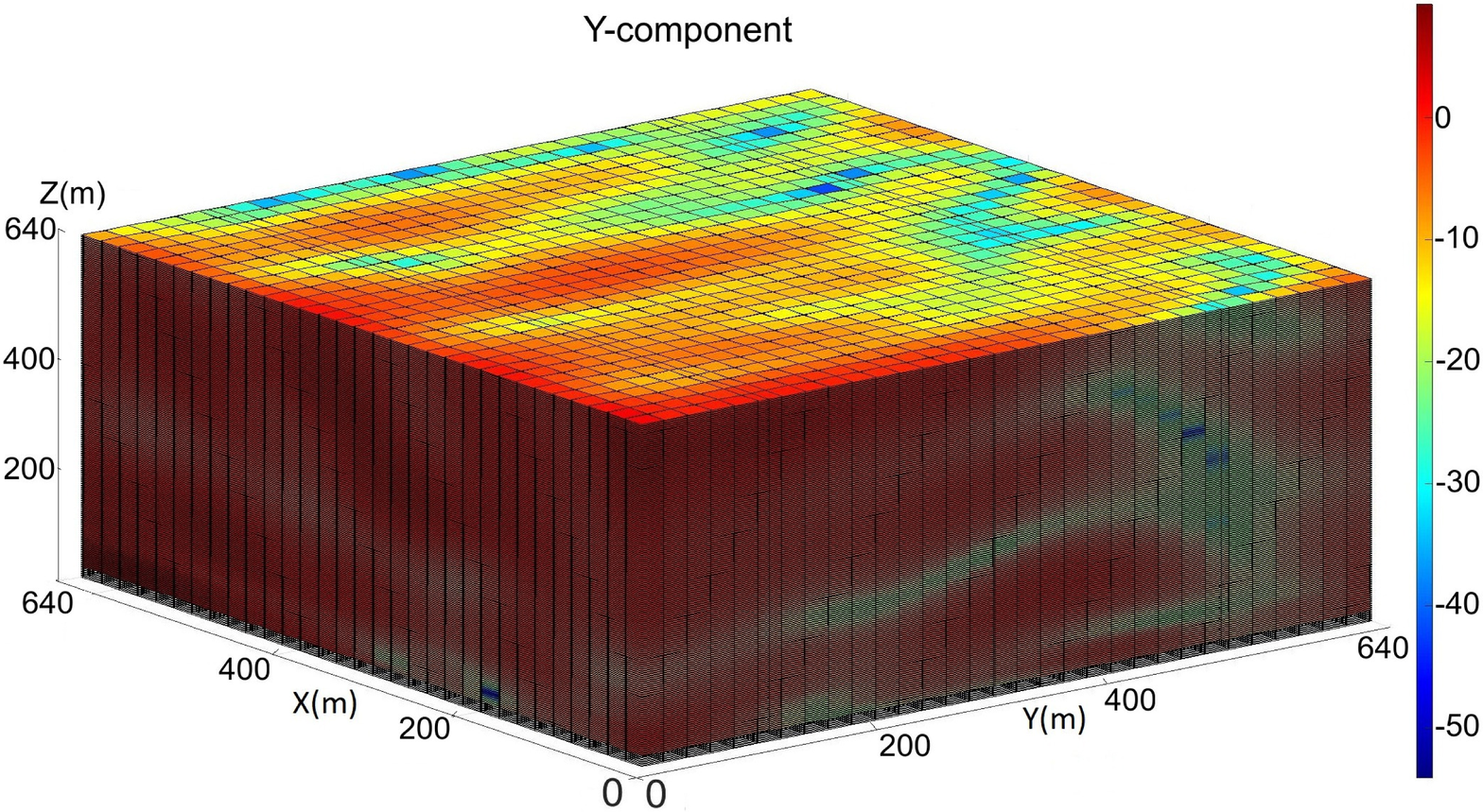}
\end{minipage}
\begin{minipage}{0.5\textwidth}
    \includegraphics[width=4.0in]{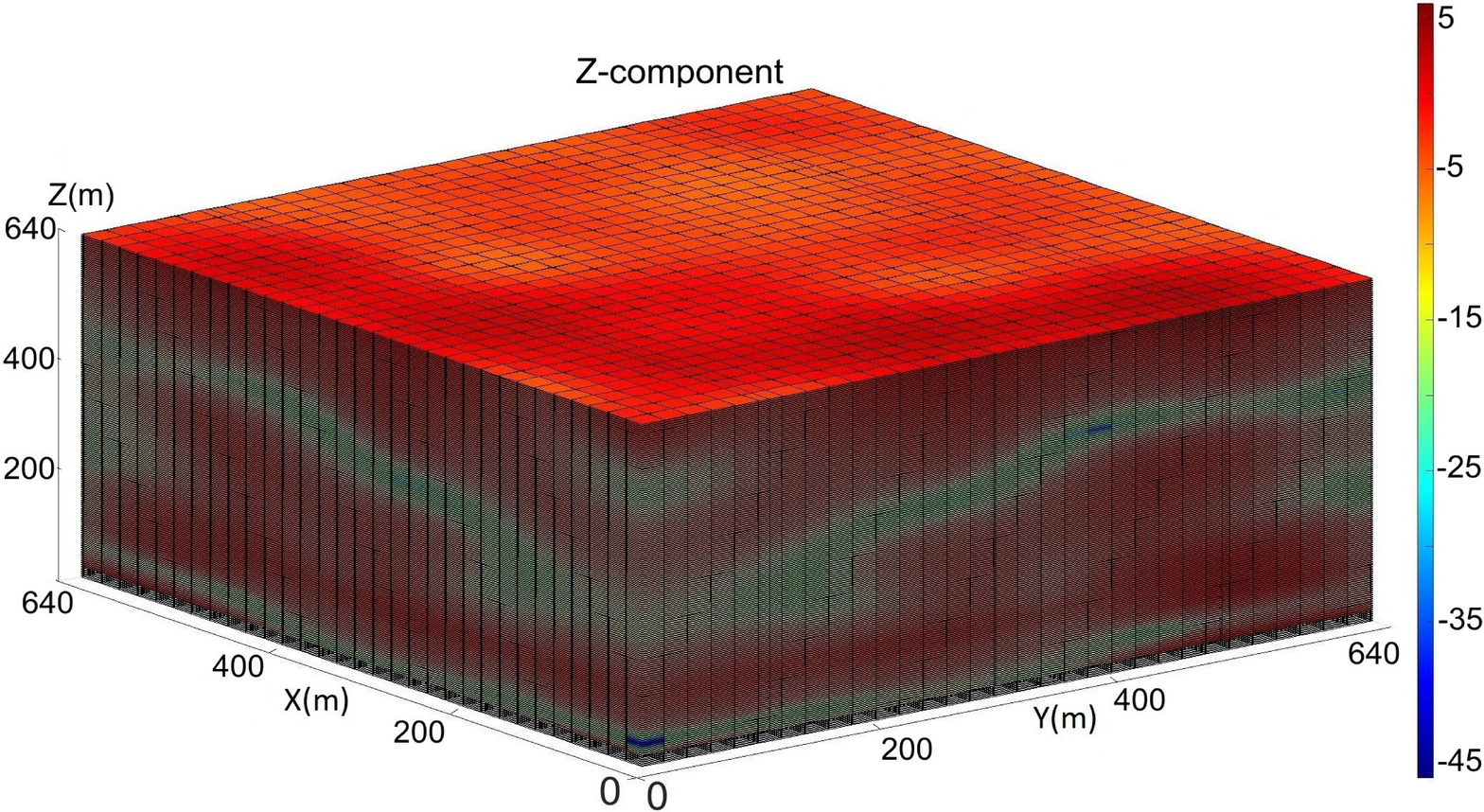}
\end{minipage}
\caption{3D view of elastic displacement components at 10 Hz in the homogeneous VTI model due to a double-couple source.}
\end{figure*}

\begin{figure*}
    \centering
    \includegraphics[width=0.7\textwidth]{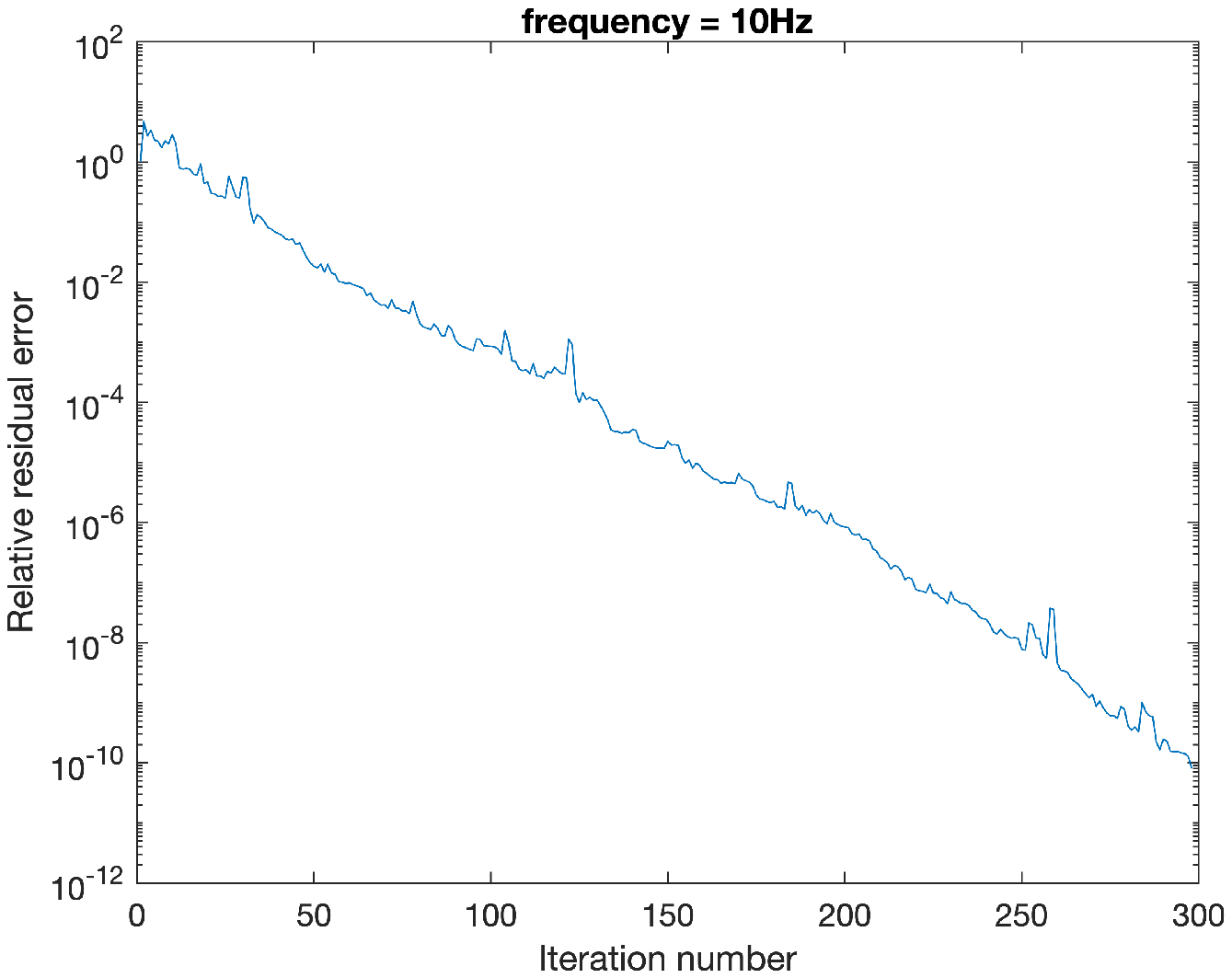}
    \caption{Plot of the relative residual error vs. iteration number for the Bi-CGSTAB solver in the case of the homogeneous VTI model.}
\end{figure*}

\begin{figure*}
\centering
\begin{minipage}{0.45\textwidth}
    \includegraphics[width=3in]{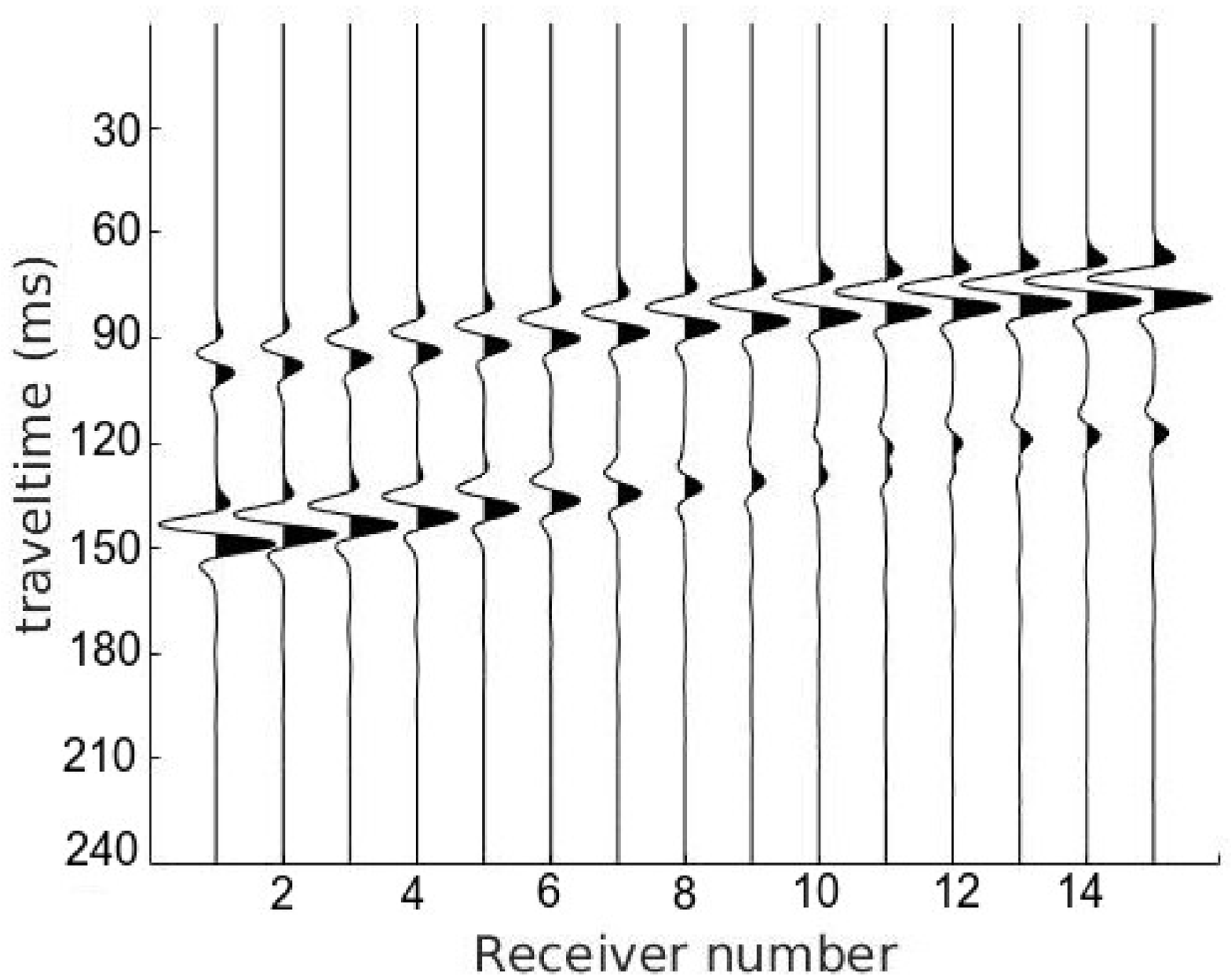}
    \centering
    \subcaption{a. Synthetic seismograms (X-component of the particle displacement) generated using the integral equation method}
\end{minipage}
\hfill
\begin{minipage}{0.45\textwidth}
    \includegraphics[width=3in]{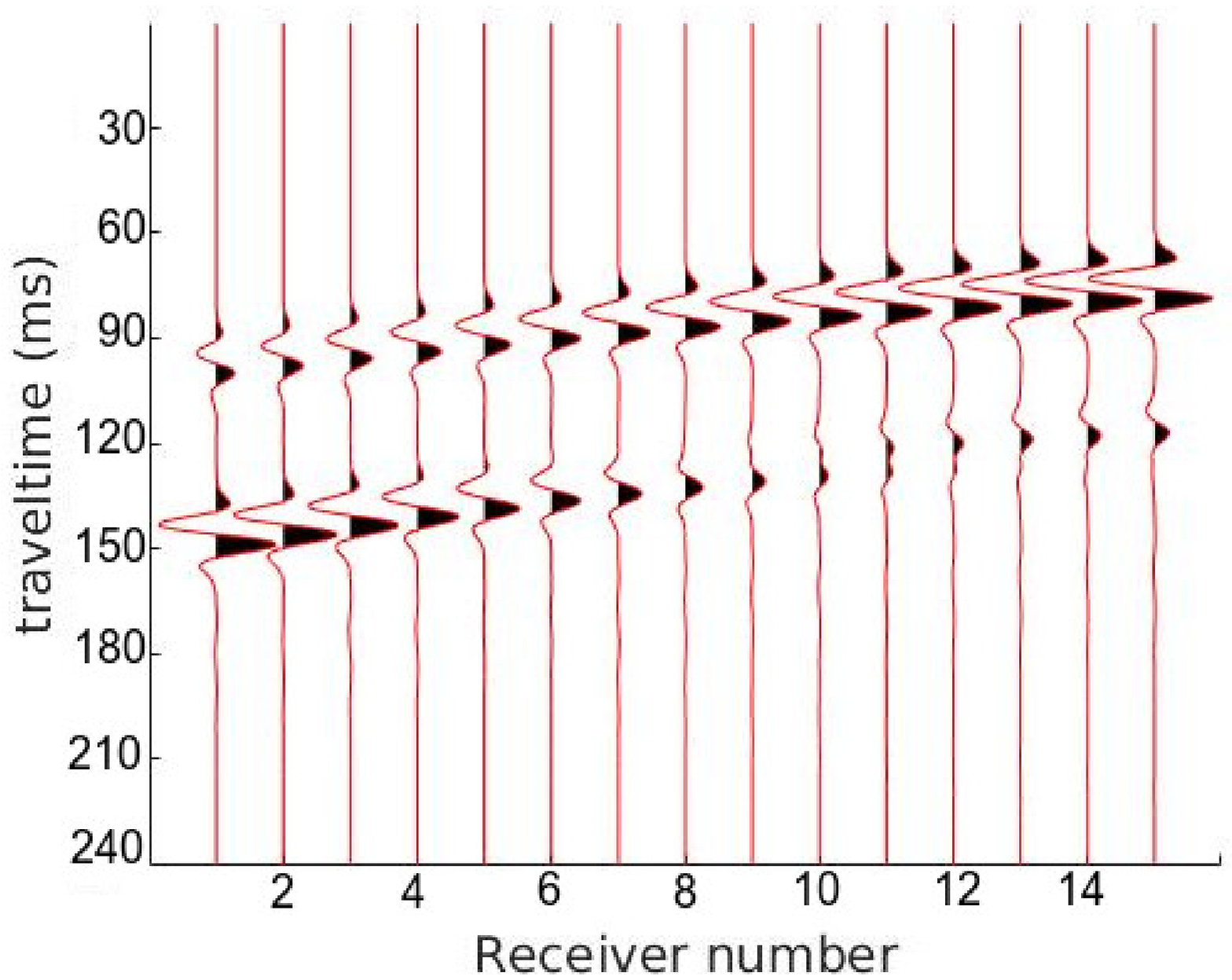}
    \centering
    \subcaption{b. Synthetic seismograms (X-component of the particle displacement) generated using the finite difference method}
\end{minipage}
\vfill
\begin{minipage}{0.45\textwidth}
    \includegraphics[width=3in]{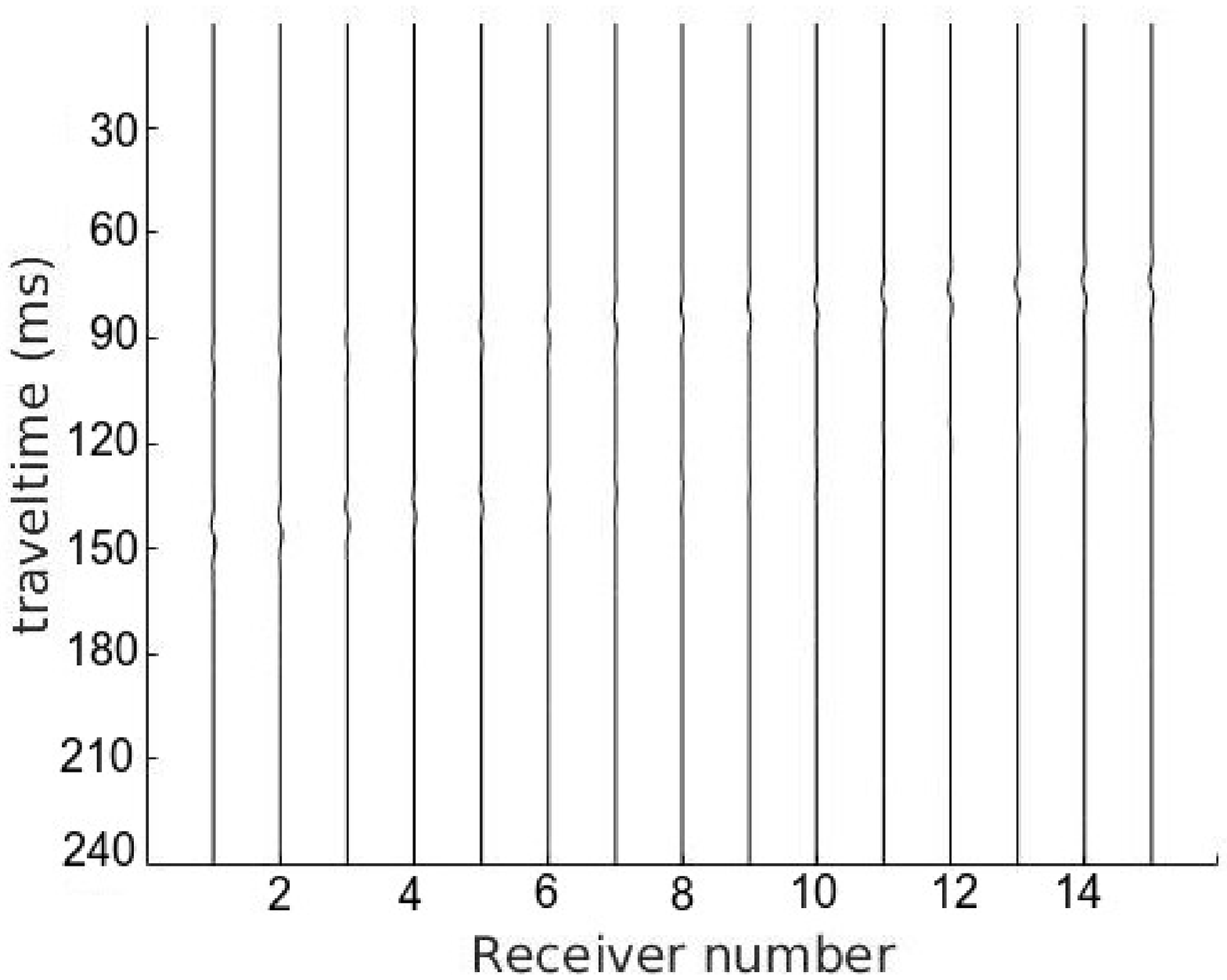}
    \centering
    \subcaption{c. Difference between the seismograms obtained using two methods}
\end{minipage}
\hfill
\begin{minipage}{0.45\textwidth}
    \includegraphics[width=3in]{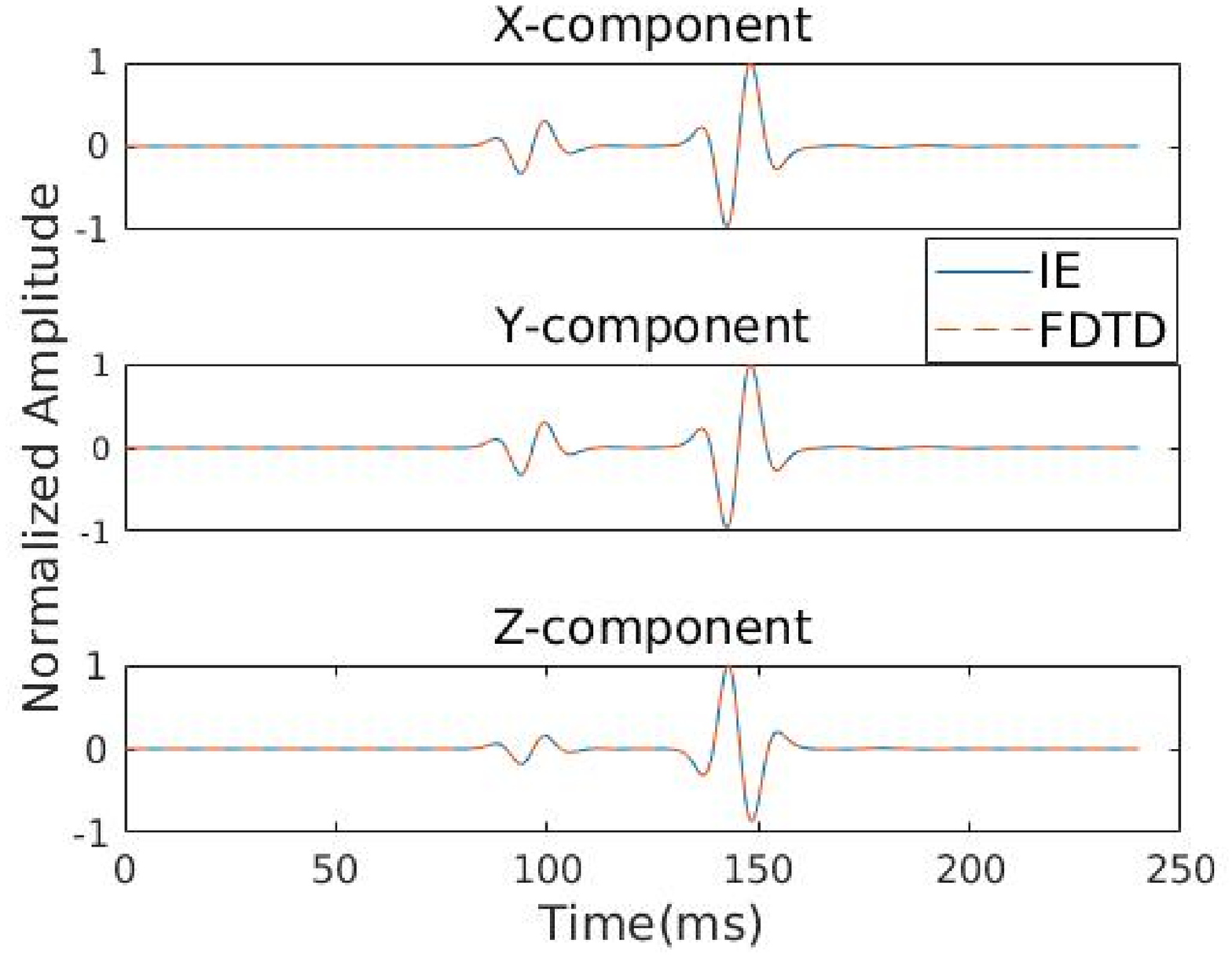}
    \centering
    \subcaption{d. Elastic displacement components at a common receiver}
\end{minipage}
\caption{Comparison between the integral equation and the finite difference methods for the model parameters given in Table 1.}
\end{figure*}

\begin{figure*}
    \centering
    \includegraphics[width=6in]{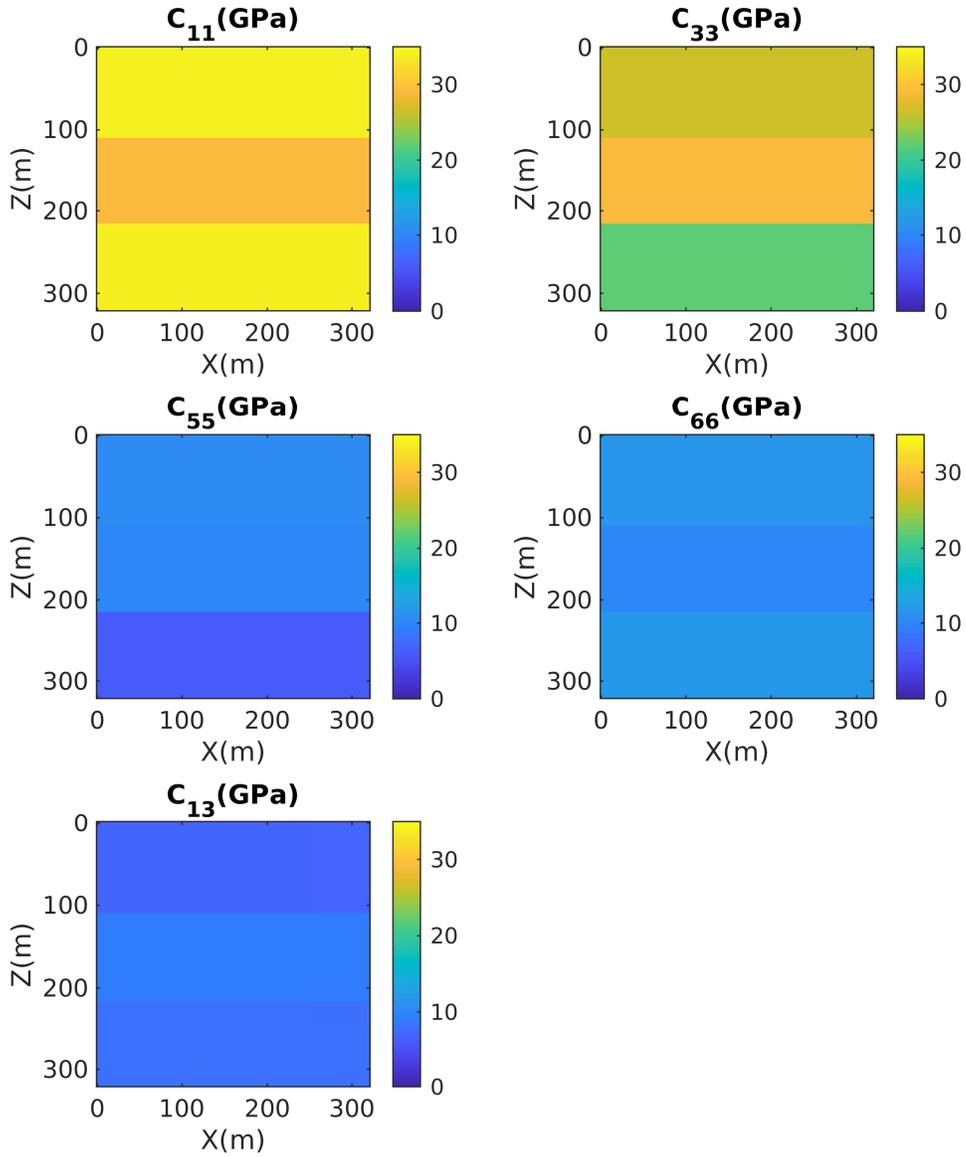}
    \caption{Model 2 - Heterogeneous VTI model.}
\end{figure*}

\begin{figure*}
\centering
\begin{minipage}{0.45\textwidth}
    \includegraphics[width=2.5in]{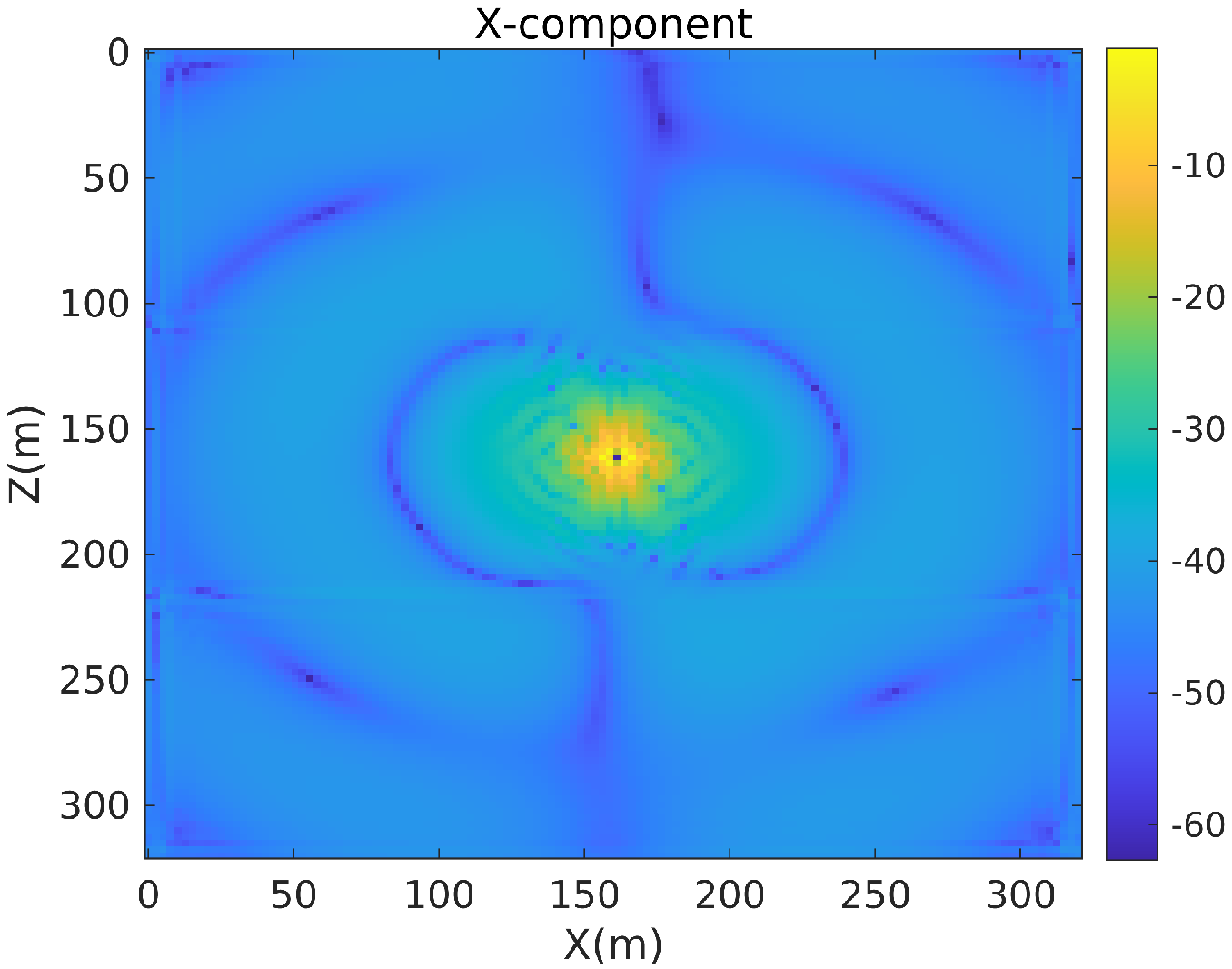}
\end{minipage}
\begin{minipage}{0.45\textwidth}
    \includegraphics[width=2.5in]{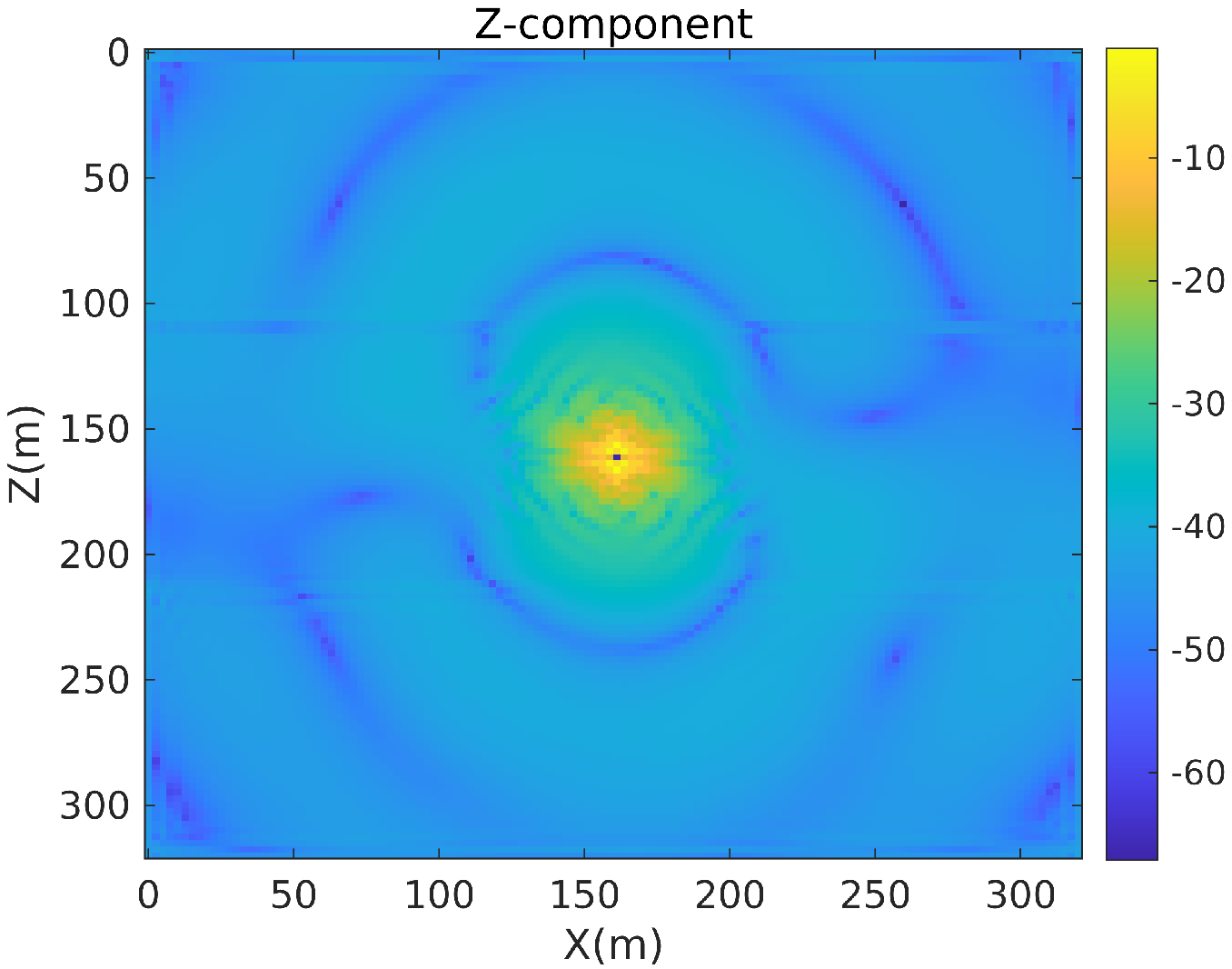}
\end{minipage}
\caption{Displacement field at 50 Hz for the heterogeneous VTI model due to an explosion-type moment tensor source.}
\end{figure*}

\begin{figure*}
    \centering
    \includegraphics[width=7in]{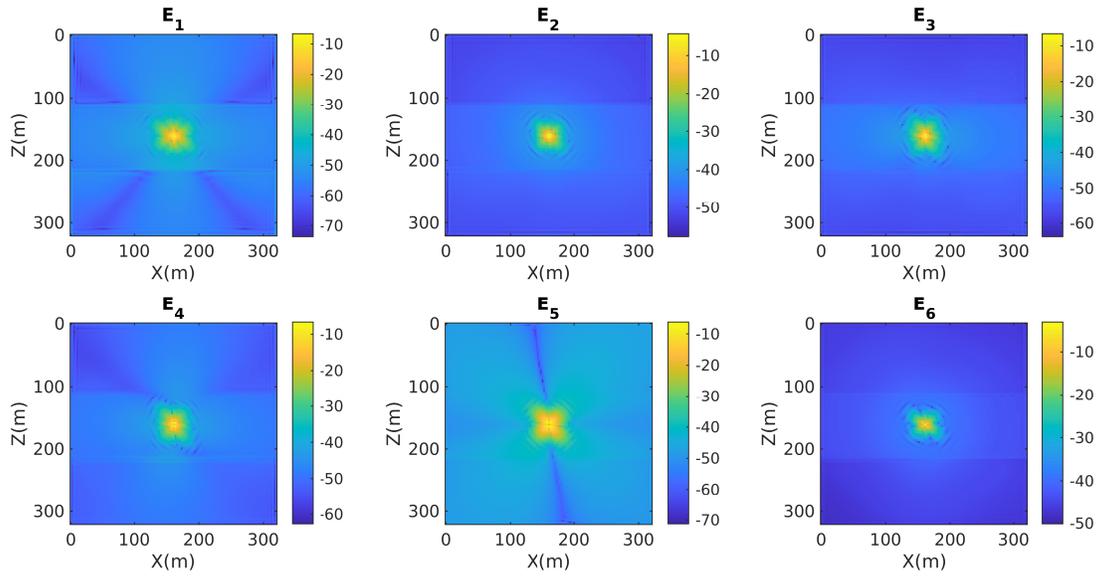}
    \caption{Strain field at 50 Hz for the heterogeneous VTI model due to an explosion-type moment tensor source.}
\end{figure*}

\begin{figure*}
    \centering
    \includegraphics[width=0.7\textwidth]{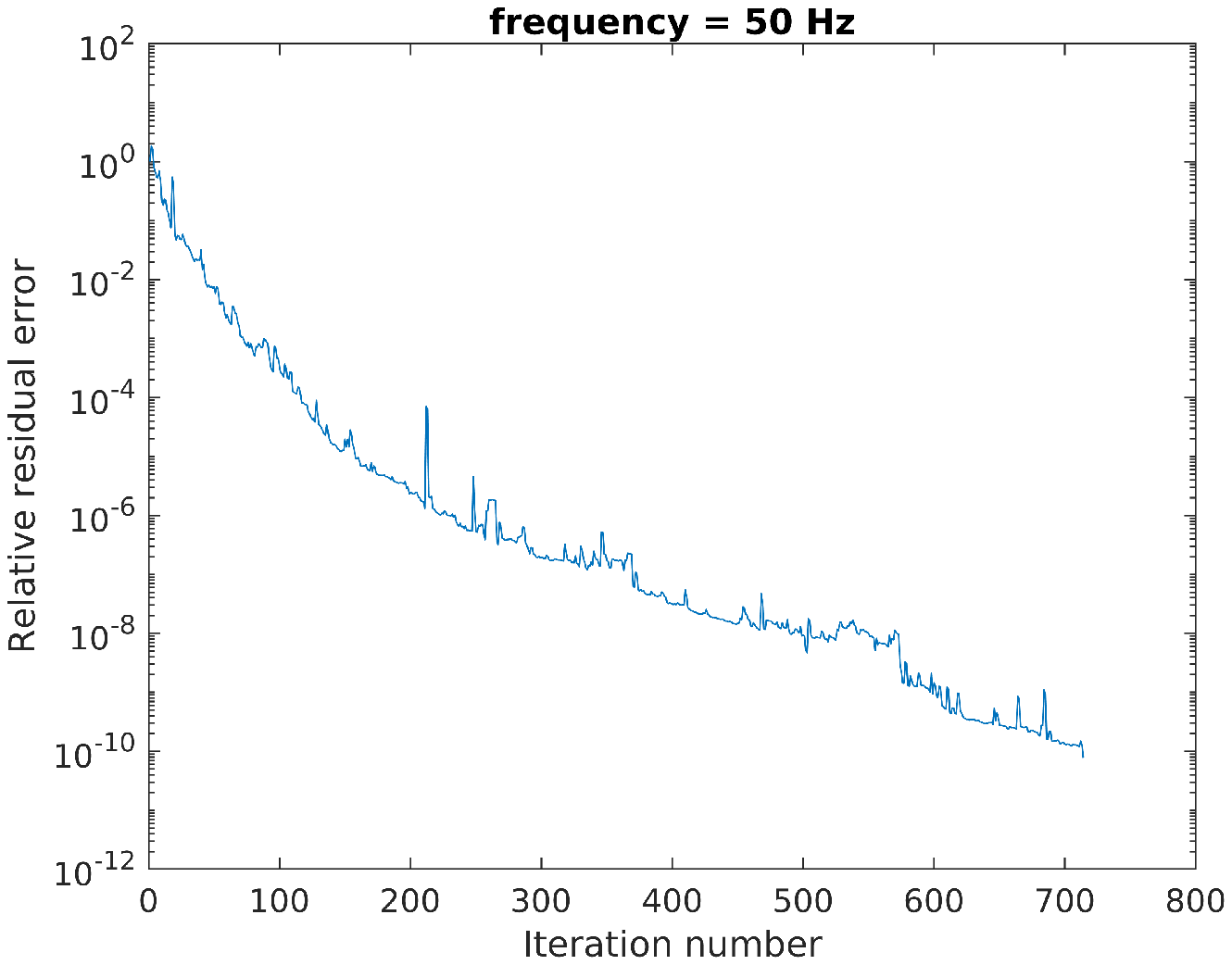}
    \caption{Plot of the relative residual error vs. iteration number for the Bi-CGSTAB solver in the case of the heterogeneous VTI model.}
\end{figure*}

\begin{figure*}
\centering
\begin{minipage}{0.45\textwidth}
    \includegraphics[width=3in]{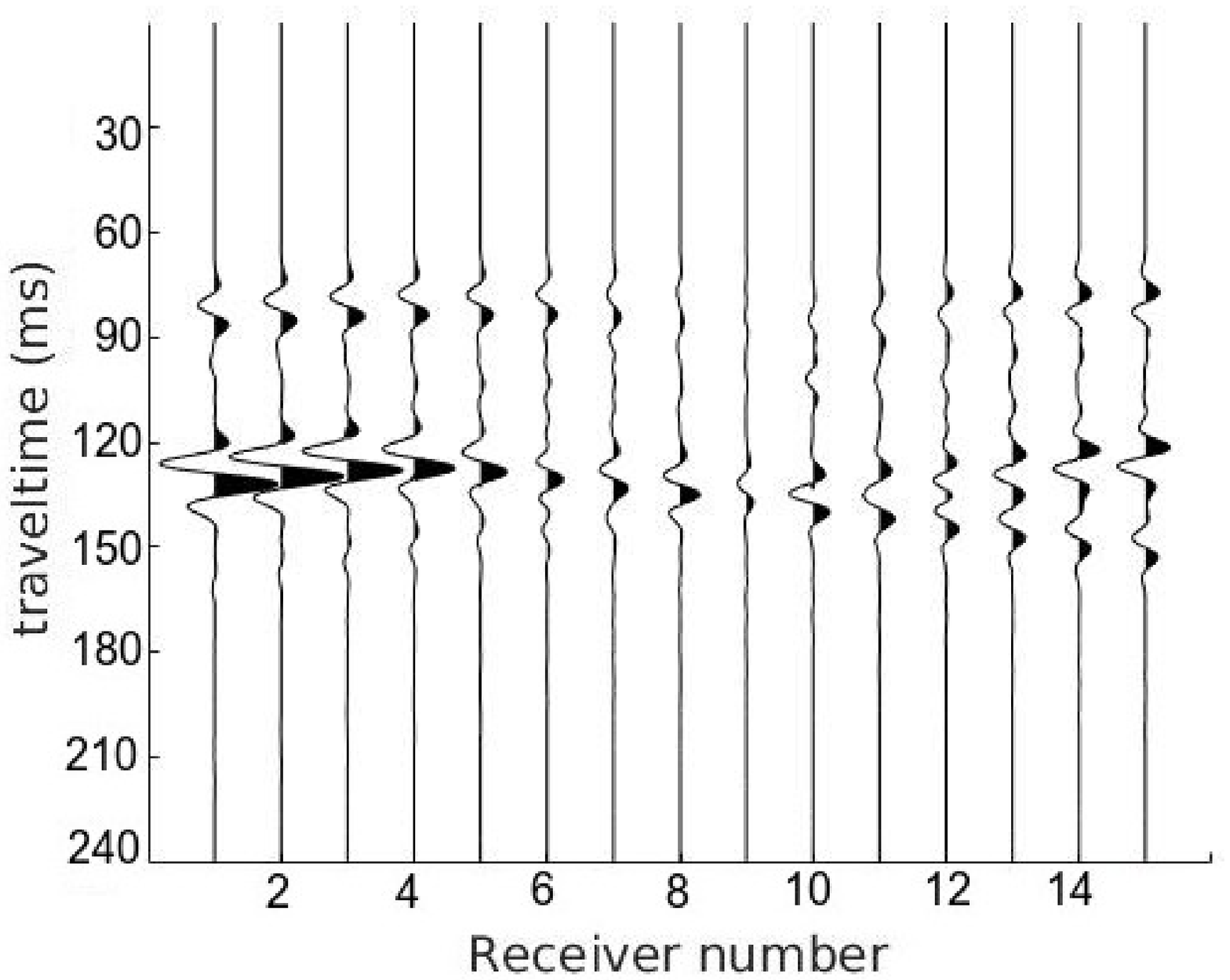}
    \centering
    \subcaption{a. Synthetic seismograms (X-component of the particle displacement) generated using the integral equation method}
\end{minipage}
\hfill
\begin{minipage}{0.45\textwidth}
    \includegraphics[width=3in]{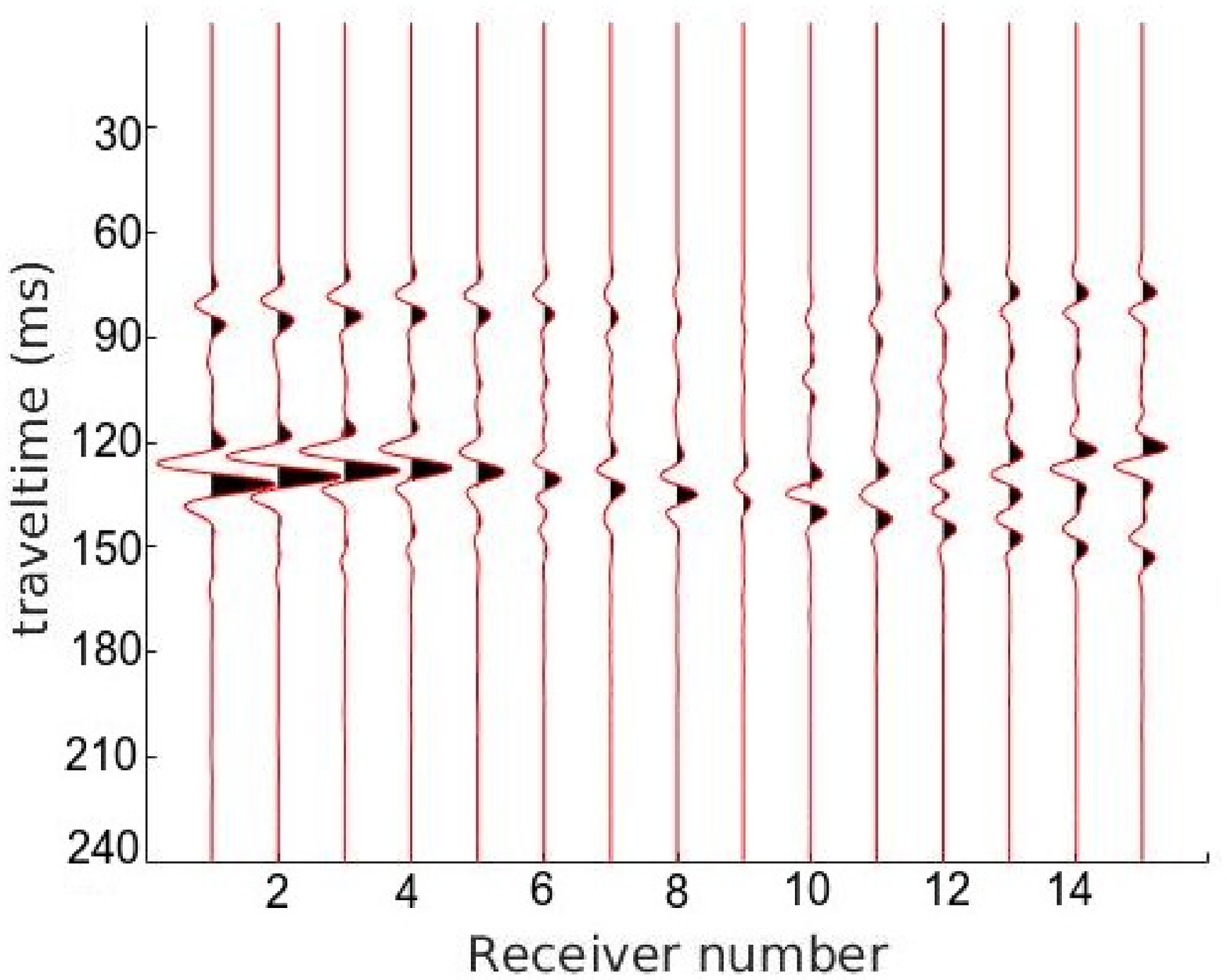}
    \centering
    \subcaption{b. Synthetic seismograms (X-component of the particle displacement) generated using the finite difference method}
\end{minipage}
\vfill
\begin{minipage}{0.45\textwidth}
    \includegraphics[width=3in]{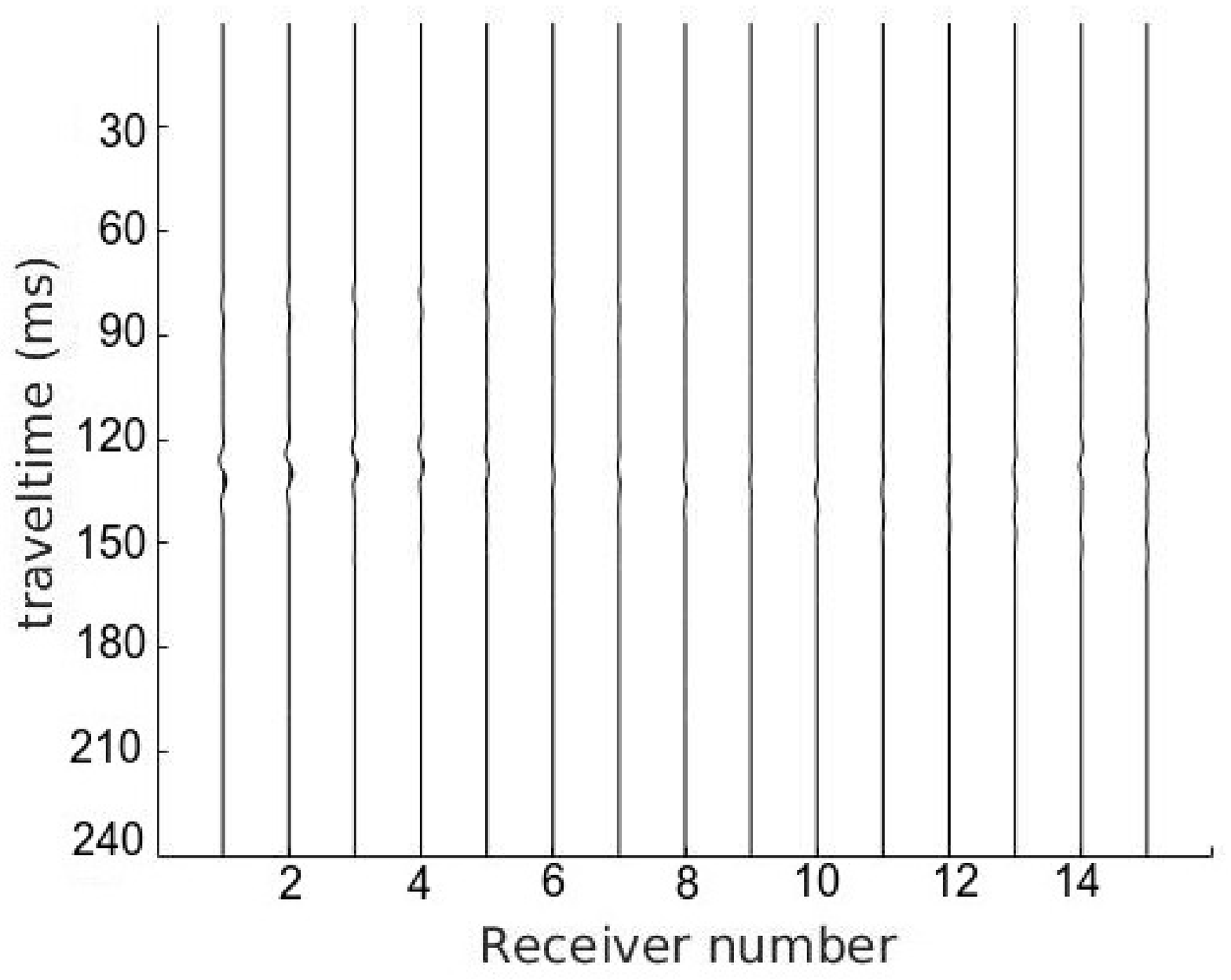}
    \centering
    \subcaption{c. Difference between the seismograms obtained using two methods}
\end{minipage}
\hfill
\begin{minipage}{0.45\textwidth}
    \includegraphics[width=3in]{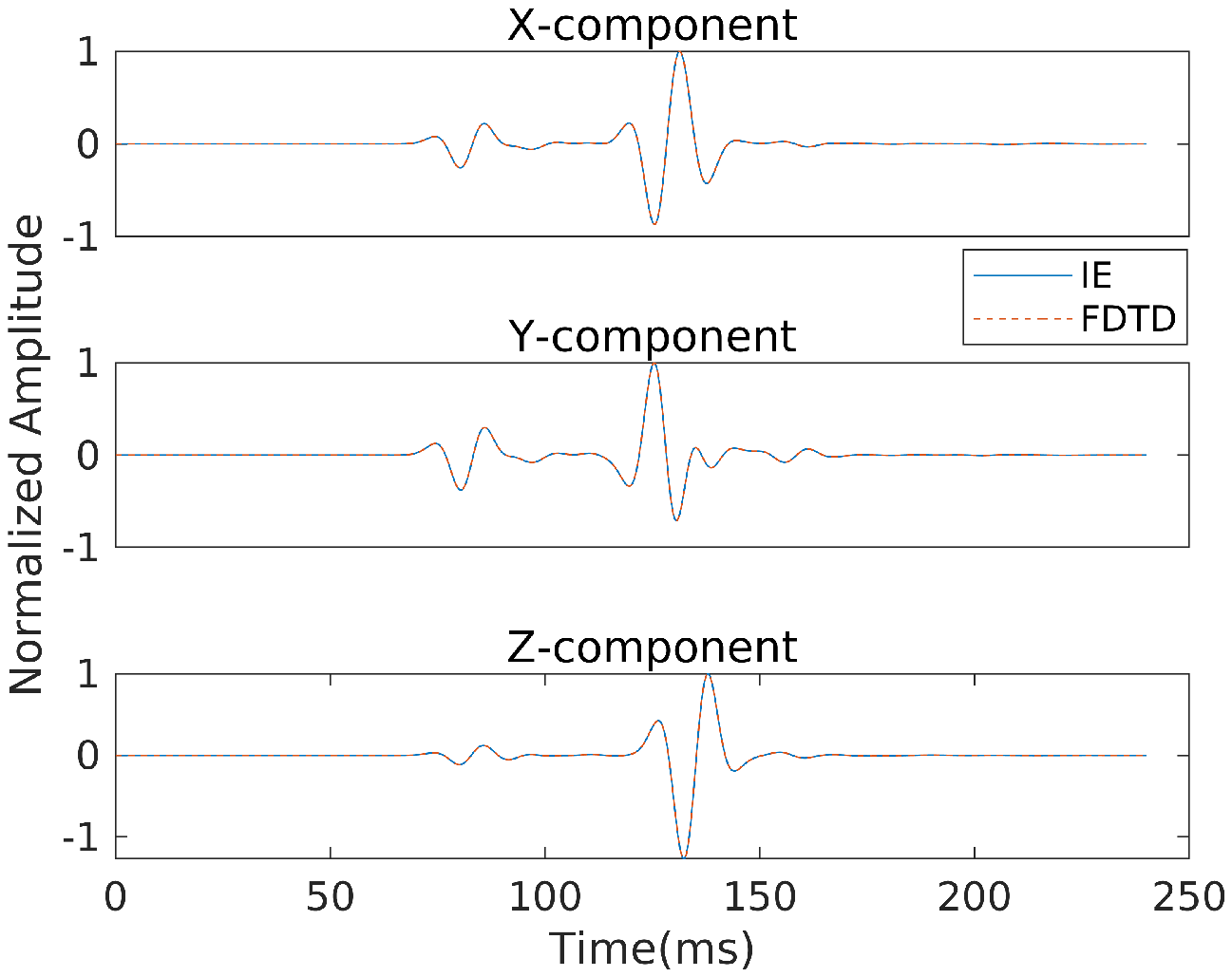}
    \centering
    \subcaption{d. Elastic displacement components at a common receiver}
\end{minipage}
\caption{Comparison between the integral equation and the finite difference methods for the model parameters given in Table 3.}
\end{figure*}

\begin{figure*}
    \centering
    \includegraphics[width=7in]{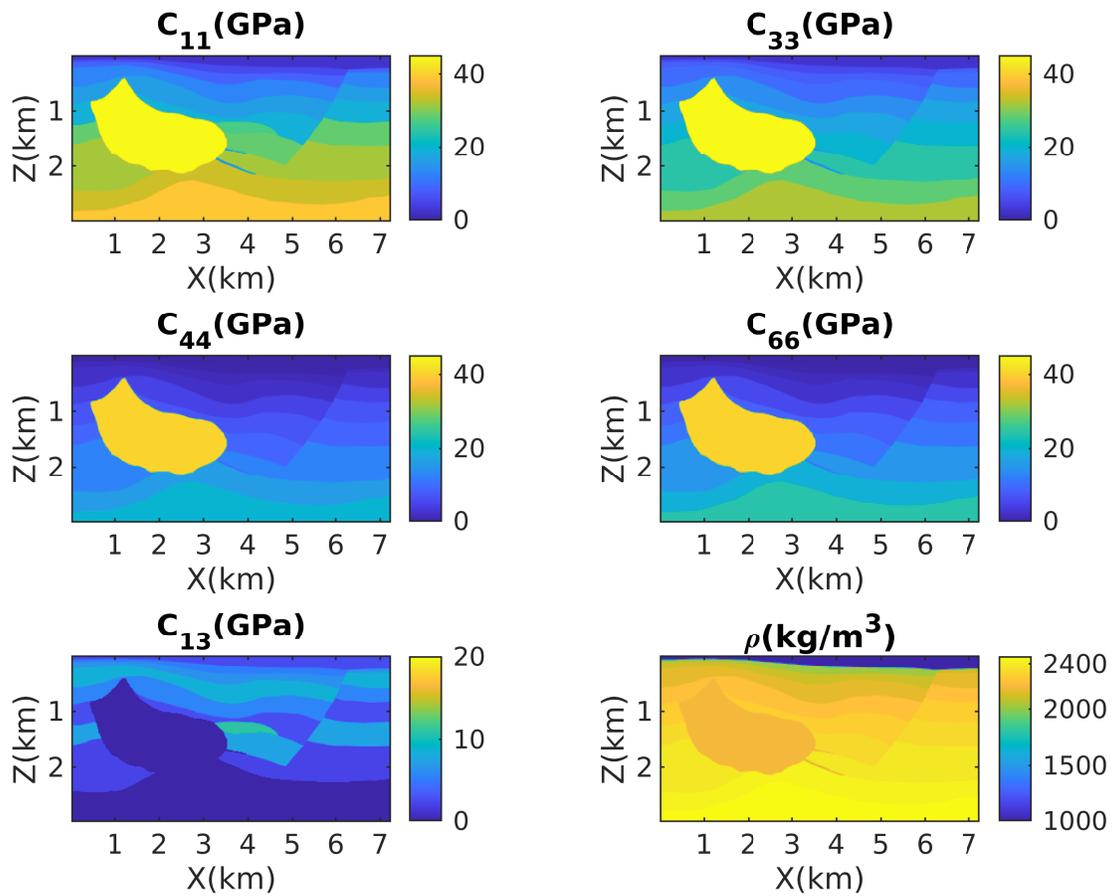}
    \caption{Model 3 - Hess model.}
\end{figure*}

\begin{figure*}
    \centering
    \includegraphics[width=1.2\textwidth]{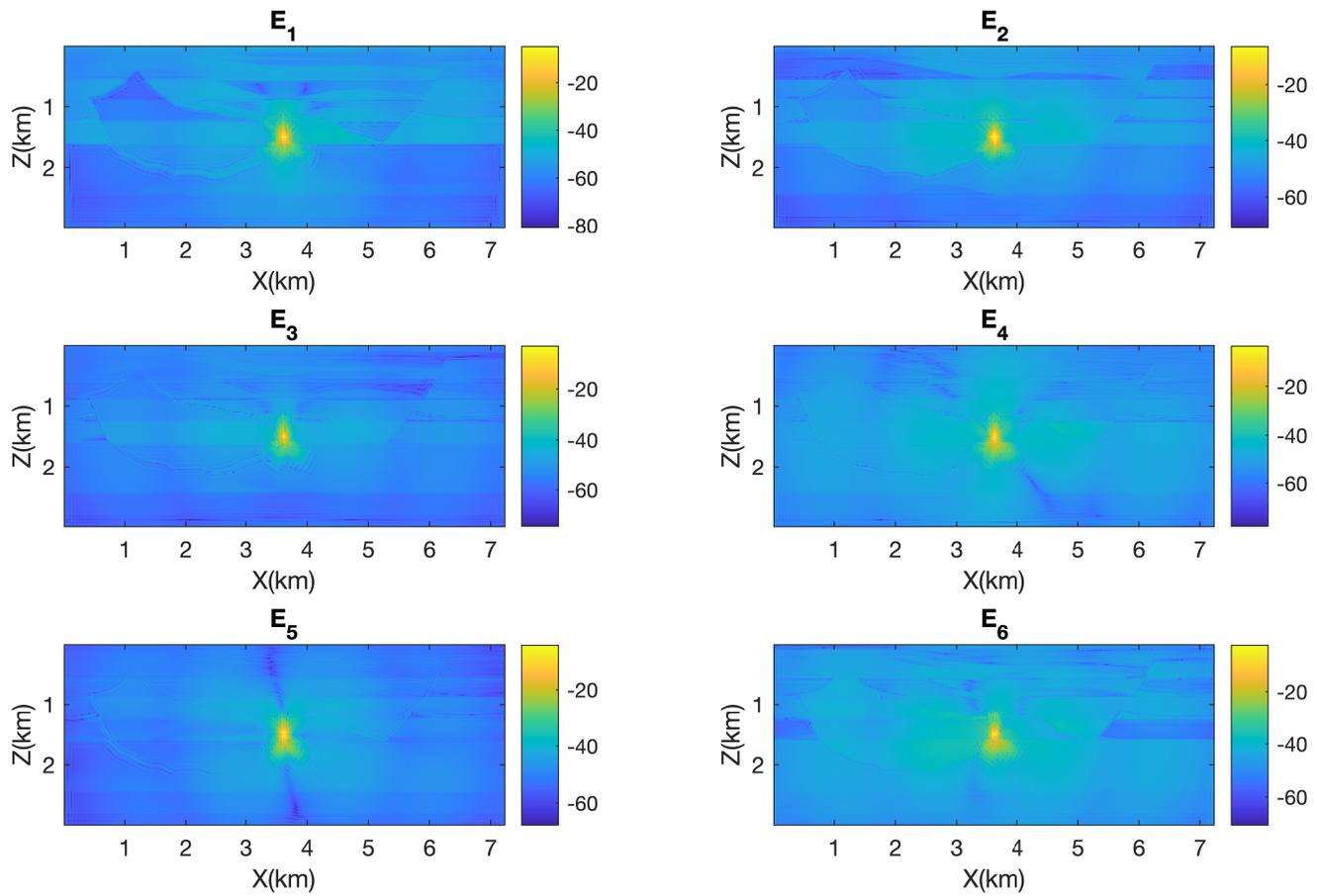}    
    \caption{Strain field due to an explosion-type moment tensor source at 10 Hz frequency for the Hess model. }
\end{figure*}

\begin{figure*}
    \centering
    \includegraphics[width=0.6\textwidth]{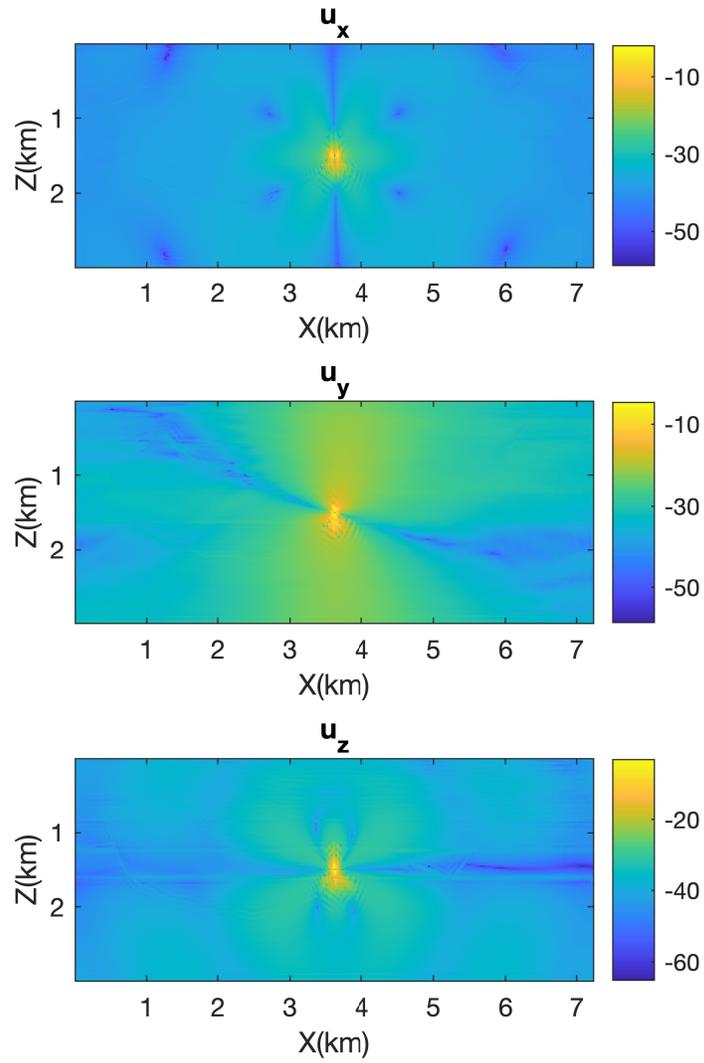}
    \caption{Displacement field due to an explosion-type moment tensor source at 10 Hz frequency for the Hess model.}
\end{figure*}

\begin{figure*}
    \centering
    \includegraphics[width=0.7\textwidth]{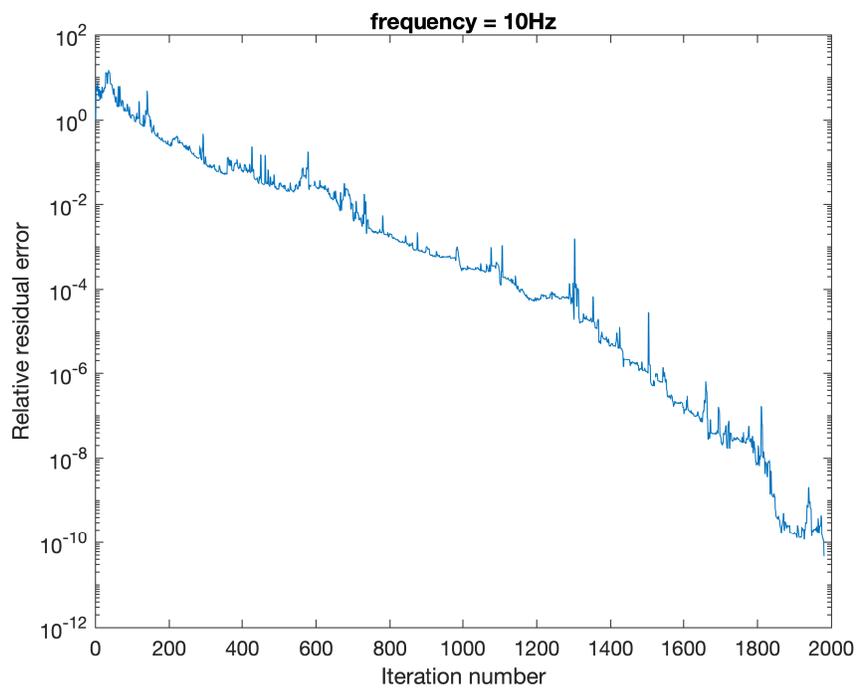}
    \caption{Plot of the relative residual error vs. iteration number for the Bi-CGSTAB solver in the case of the variable density Hess model.}
\end{figure*}

\label{lastpage}
\end{document}